\documentclass[12pt,a4paper]{article}
\usepackage{graphicx,amssymb,amsmath}
\usepackage{graphicx, array, epsfig, subfigure}
\usepackage{multirow}
\usepackage{color}
\usepackage{xcolor}

\usepackage{subfigure}  % use for side-by-side figures
\newcommand{\be}{\begin{equation}}
\newcommand{\ee}{\end{equation}}
\newcommand{\bea}{\begin{eqnarray}}
\newcommand{\eea}{\end{eqnarray}}

\catcode`@=12

\begin{document}
\begin{center}
{\bf The Violation of Equivalence Principle and Four Neutrino Oscillations 
for Long Baseline Neutrinos}\\
%{\bf Mass and Life Time of Heavy Dark Matter Decaying into IceCube PeV 
%Neutrinos}\\
\vspace{1cm}
{{\bf Madhurima Pandey$^{a,b}$} \footnote {email: madhurima0810@gmail.com},
{\bf Debasish Majumdar$^{a}$} \footnote {email: debasish.majumdar@saha.ac.in}},\\
{\normalsize \it $^a$Theory Division, Saha Institute of Nuclear Physics, HBNI}  \\
{\normalsize \it 1/AF Bidhannagar, Kolkata 700064, India } \\
\vspace{0.15cm}
{\normalsize \it $^b$ Department of Physics, School of Applied Sciences and Humanities,} \\
{\normalsize \it Haldia Institute of Technology, Haldia, West Bengal, 721657, India.} \\
\vspace{0.25cm}
{\bf Amit Dutta Banik$^{c}$} \footnote{email: amitdbanik@mail.ccnu.edu.cn}\\
{\normalsize \it $^{c}$Key Laboratory of Quark and Lepton Physics (MoE) and Institute of Particle 
Physics, Central China Normal University, Wuhan 430079, China} \\
%{\normalsize \it 152, Luoyu Avenue, Wuhan - 430079, Hubei, China}  \\
\vspace{0.25cm}
{\bf Ashadul Halder$^{d}$} \footnote{email: ashadul.halder@gmail.com}\\
{\normalsize \it $^{d}$Department of Physics, St. Xavier's College,} \\
{\normalsize \it 30, Mother Teresa Sarani, Kolkata - 700016, India}  \\
\vspace{1cm}
\end{center}

\begin{center}
{\bf Abstract}
\end{center}

{\small
Violation of equivalence principle predicts that neutrinos of different 
flavour couple differently with gravity. Such a scenario can give rise to 
gravity induced flavour oscillations in addition 
to the usual mass flavour neutrino oscillations during the neutrino 
propagation. Even if the equivalence principle is indeed violated, their 
measure will be extremely small. We 
explore the possibility to probe the violation of equivalence principle (VEP) 
for the case of long baseline (LBL) neutrinos in a 4-flavour neutrino framework 
(3 active + 1 sterile) where both mass and gravity induced oscillations are 
considered. To this end, we have explicitly calculated the oscillation 
probability in 4-flavour framework that includes in addition to the 
mass-flavour mixing in matter, the gravity-flavour mixing also. The energy 
eigenvalues are then obtained by diagonalising such a 4-flavour mixing 
matrix. The formalism is then employed to estimate the wrong and right 
sign muon yields at a far detector for neutrinos produced in a neutrino 
factory and travel through the Earth matter. These results are compared 
with the similar
estimations when the ususal three active neutrinos are considered.

%For the muon neutrino flux from a neutrino factory or accelerator, 
%the muon yields are computed and compared at a far end detector with and 
%without VEP. For the first time, we explore a possibility demonstrating that 
%in 4-neutrino (3 active + 1 sterile) framework considering the neutrino 
%oscillations with both VEP and matter effect, even a very small
%violation of equivalence principle can be probed in a long 
%baseline neutrino experiment. {\bf These results are compared with the similar 
%estimations when the ususal three active neutrinos are considered.}}
\newpage
\section{Introduction}
The oscillatory nature of neutrinos \cite{bil} from one type of flavour to another is now well established by various terrestrial experiments with neutrinos 
having natural origin such as atmospheric, solar and man made such as reactor 
\cite{reactorano,reactorano1,reactorano2} or accelerator neutrinos. The mass 
eigenstates and the weak interaction eigenstates of neutrinos not being the 
same, neutrino flavour eigenstate in a coherent neutrino beam can oscillate 
into a eigenstate having different flavour after traversing a distance. These 
oscillations occur due to the phase difference that is acquired by a neutrino 
eigenstate during its propagation and this phase difference depends on the 
baseline length and the mass square difference of two neutrino mass 
eigenstates. The massive nature of neutrinos is 
established by experimental phenomenon of the oscillations. The framework of 
Standard Model (SM) of particle physics does not have mechanisms to explain how the neutrinos acquire masses and theories beyond the SM framework needs to be 
invoked for explaining the neutrino mass.

In association to the neutrino mass, violation of the equivalence principle 
(VEP) \cite{gas,gas1} can also induce neutrino oscillations. If the equivalence 
principle is indeed violated in nature, different types of neutrinos couple
differently with gravity which means that different neutrino flavour 
eigenstates interact with the gravitation field with different strengths. Thus 
in this situation the gravitational coupling (constant) $G$ is different for 
different types of neutrinos. This leads to the
fact that the gravity eigenstates of neutrinos are not identical as those of 
their weak interaction eigenstates. An important feature of Einstein's 
general theory of relativity is the equivalence principle which affirms that 
the inertial mass and the gravitational mass are the same. This is 
stated that an observer standing on the Earth experiences the gravitational 
force which is same as the pseudo force experienced by the same observer in 
accelerated reference frame. Therefore, if the equivalence principle is indeed 
violated, then the coupling strengths of neutrinos with the gravitational field 
as well as the gravitational constant ($G$) are no more universal.

A general consequence of VEP is the gravitational redshift - while 
propagating through a gravitational field the energy $E$ of a neutrino
will be shifted by an amount $E' = \sqrt{g_{00}} E = E (1 - \displaystyle 
\frac {GM} {R} ) = E (1 + \phi)$ \cite{gas1,gas2}, where the gravitational 
potential ($\phi$) \cite{pot}
is defined as $\phi = \frac {GM} {R}$, $M$ and $R$ being the mass of the 
source and the distance over which the gravitational field operates respectively 
\footnote{In the presence of gravitational field, the proper time in a 
curve manifold is $d\tau = \sqrt{g_{\mu \nu} dx^{\mu}dx^{\nu}}$ which can 
lead to the relation  $E' = \sqrt{g_{00}} E$. The proper time ($d\tau$) 
relates to the coordinate time ($dt$) through $d\tau = \sqrt{g_{\mu \nu} 
dx^{\mu}dx^{\nu}}$ (when clock is at rest). If a distant star is emitting $N$ 
number of waves having frequency $f_{\rm star}$ and proper time interval 
$\Delta {\tau_{\rm star}}$ 
respectively and if the Earth is detecting the same with 
frequency $f_{\rm Earth}$ and proper time interval $\Delta {\tau_{\rm Earth}}$, 
then $\displaystyle\frac {f_{\rm star}} {f_{\rm Earth}} = 
\displaystyle\frac {\Delta {\tau_{\rm Earth}}} {\Delta {\tau_{\rm star}}} = 
\displaystyle\frac {\sqrt{g_{00} (x_{\rm Earth})}} 
{\sqrt{g_{00} (x_{\rm star})}} = \sqrt{\left ( \displaystyle\frac 
{1+2 \phi_{\rm Earth}} {1+2 \phi_{\rm star}} \right )} = 1 + |\Delta {\phi|}$.}. 
By virtue of the equivalence principle, energies of different types of 
neutrinos are shifted by an equal amount and eventually while they traverse
through the gravitational field, the phase difference between two types of 
neutrinos may not be generated in this case. But if the violation of equivalence 
principle is violated then for different types of neutrinos 
the energies will be shifted differently. The VEP will induce a 
phase $\sim \Delta {E} L$, $\Delta {E} = |E_i - E_j|$, $E_i$ and $E_j$ being the 
red-shifted energies of the neutrino species $i$ and $j$ respectively and 
$L$ defines the baseline length from the source to the Earth through which 
neutrino propagates. $E_i, E_j$ are the energy eigenstates in gravity basis. 
Similar to the case for mass-flavour oscillations, the acquired phase difference will 
generate a gravity induced oscillations between different 
flavours of neutrinos having the oscillatory part 
$\sim |\Delta E L| = |\Delta f_{ij}| LE$, where 
$|\Delta f_{ij}| = |f_i - f_j|$, $f_i$ is defined as 
$f_i = \frac{G_i M} {R} = (\frac{GM} {R}) \alpha_i = \phi \alpha_i$, $G_i$ 
being the gravitational coupling of the neutrino having index $i$.

In this work, we study the effects of violation of equivalence principle with 
three active and one sterile neutrino \footnote{In a previous work, velocity induced 
oscillations in matter had been addressed \cite{amit} in case of long baseline neutrinos similar to
the present analysis. But this work was performed in the context of only three active
neutrinos.}. To this end, we first obtain the evolution equation 
of the neutrino including three effects namely the mass-flavour mixing, 
the gravity-flavour mixing and the matter effect. The $(4 \times 4)$ 
evolution matrix is then diagonalised by proper unitary transformation and 
the energy eigenvalues are obtained explicitly. This enables to compute 
the phase difference between any two neutrinos during its propagation and 
hence the expressions for different oscillation probabilities with the 
oscillatory part have been written.
Using the latest experimental limits on active-sterile neutrino mixing and mass square difference $\Delta m_{41}^2$ for 
normal hierarchy of neutrino mass eigenstates (and assumed normal 
hierarchy for neutrino gravity eigenstates) and best fit values of active 
neutrino mixing parameters (mixing angles and mass square 
differences), we obtain new four flavour gravity induced neutrino oscillation 
probabilities which also include the matter effect induced by the matter through which the neutrinos travel.
We show that VEP will induce new set of parameters $\Delta f_{ij}$ which 
change the neutrino oscillation probabilities significantly.
%even if $\Delta f_{ij}$ assumes a value as small as $\sim 10^{-24}$. 
We demonstrate this in case of a neutrino beam propagating 
through a baseline of 7000 km inside Earth matter. 
Therefore, even if VEP is very small, it will significantly affect  
the number of muon yields (from $\nu_\mu$) at the far detector in a long 
baseline (LBL) neutrino experiment.
In this work, we compute our results of these neutrino yields considering an 
LBL neutrino experiment for a baseline length 
of around 7000 km with the end detector to be a iron calorimeter (ICAL) 
of 50 kTon such as the one proposed for the India-based Neutrino Observatory or INO \cite{Kumar:2017sdq} and at the origin of the neutrino source is from a 
neutrino factory or accelerator such as CERN. 
We calculate the number of right sign and wrong sign 
(explained later) muon yields at the end detector and their variations with 
the change in $\Delta f_{ij}$ (the VEP effect). We also mention that perhaps 
this is for the first time, neutrino oscillations 
in 4-flavour scenario has been worked out in detail and expressions for oscillation
probabilities are obtained incorporating both gravity induced and matter induced mass flavour oscillations.

We organise the paper in the following manner. In Section 2 we present a 
brief discussion about the formalism of gravity induced as well as mass 
induced oscillations in matter within a 4-flavour framework. The calculational results are 
furnished in Section 3 which is divided into two subsections. In Section 
3.1 we discuss about how the oscillation probabilities vary with the gravity 
effects, while Section 3.2 deals with the possible neutrino 
induced muon yield in long baseline (LBL) experiments in the presence of both  
gravity induced oscillations and mass induced oscillations in matter. Finally in 
Section 4 the paper is summarized with some discussions. 

\section{Neutrino Oscillations in Matter with VEP in 4-flavour Scenario}
Neutrino oscillations would arise because of the non zero nature of neutrino 
masses. The essence of this phenomena was first observed by Pontecorvo 
\cite{pont,pont1} in 1957, while Maki, Nakagawa and Sakata (1962) first pointed 
out the possibility of the arbitrary mixing between the two massive neutrino 
states.

In the case of massive neutrinos, the neutrino flavour eigenstates 
$|\nu_\alpha \rangle$ produced in a charged current weak interactions 
can be expressed as the linear combination of the mass eigenstates 
$|\nu_i \rangle$ via a unitary mixing matrix $U$ (with matrix elements denoted as $U^*_{\alpha i}$).
\bea
| \nu_\alpha \rangle &=& \displaystyle\sum_{i=1}^{n} U^*_{\alpha i} 
|\nu_i \rangle\,\, ,
\label{form1}
\eea
where the number of neutrino species is indicated by $n$. In what follows we consider the mixing matrix $U$ to the real (no CP violating phases) and for a 4-flavour scenario $U\equiv U_{(4\times 4)}$. In this work we 
consider an extra sterile neutrino ($\nu_s$) in addition to the three active 
neutrinos ($\nu_e, \nu_\mu, \nu_\tau$). For this 4-flavour 
(3 active + 1 sterile) scenario, the relation between the neutrino flavour 
eigenstates and the mass eigenstates can be parameterized as 
\bea
\left (\begin{array}{c}
\nu_{e} \\
\nu_{\mu} \\
\nu_{\tau} \\
\nu_{s} \end{array} \right )
= U_{(4 \times 4)} \left (\begin{array}{c}
\nu_{1} \\
\nu_{1} \\
\nu_{3} \\
\nu_{4} \end{array} \right )\,\,
%\eea
%\bea
= \left ( \begin{array}{cccc}
U_{e1} & U_{e2} & U_{e3} & U_{e4} \\
U_{\mu1} & U_{\mu2} & U_{\mu3} & U_{\mu4} \\
U_{\tau1} & U_{\tau2} & U_{\tau3} & U_{\tau4} \\
U_{s1} & U_{s2} & U_{s3} & U_{s4} \end{array}
 \right )
\left (\begin{array}{c}
\nu_{1} \\
\nu_{1} \\
\nu_{3} \\
\nu_{4} \end{array} \right )\,\,  ,
\label{form2}
\eea
where $U_{\alpha i}$ etc. are the elements of the Pontecorvo - Maki - 
Nakagawa - Sakata (PMNS) matrix $U_{(4 \times 4)}$ \cite{pmns}.

The PMNS matrix $U_{(4 \times 4)}$ depends on the mixing angles and CP 
violating phases. In this work, we assume CP conservation in the neutrino sector. The 
PMNS matrix can now be described in terms of the successive rotations ($R$), 
which are functions of the six mixing angles $\theta_{14},\,\theta_{24},\,\theta_{34},\,\theta_{13},\,\theta_{12},\,\theta_{23}$ in 4-flavour scenario 
\cite{element,madhu1}.
\bea
U_{(4 \times 4)} &=& R_{34}(\theta_{34})R_{24}(\theta_{24})R_{14}(\theta_{14})
R_{23}(\theta_{23})R_{13}(\theta_{13})R_{12}(\theta_{12})\,\,  ,
\label{form4}
\eea
where these six orthogonal matrices ($R$) can be written as 
{\small
\bea
R_{34} (\theta_{34}) &=& \left (\begin{array}{cccc}
1 & 0 & 0 & 0 \\
0 & 1 & 0 & 0 \\
0 & 0 & c_{34} & s_{34} \\
0 & 0 & -s_{34} & c_{34} \end{array} \right )\,\, ,
R_{24} (\theta_{24})  =  \left (\begin{array}{cccc}
1 & 0 & 0 & 0 \\
0 & c_{24} & 0 & s_{24} \\
0 & 0 & 1 & 0 \\
0 & -s_{24} & 0 & c_{24} \end{array} \right )\, , \nonumber \\
R_{14} (\theta_{14}) &=&  \left (\begin{array}{cccc}
c_{14} & 0 & 0 & s_{14} \\
0 & 1 & 0 & 0 \\
%0 & 0 & 1 & 0 \\
0 & 0 & 1 & 0 \\
-s_{14} & 0 & 0 & c_{14} \end{array} \right )\, ,
R_{12} (\theta_{12})  =  \left (\begin{array}{cccc}
c_{12} & s_{12} & 0 & 0 \\
-s_{12} & c_{12} & 0 & 0\\
0 & 0 & 1 & 0 \\
0 & 0 & 0 & 1\end{array} \right )\, , \nonumber \\
R_{13} (\theta_{13})  &=&  \left (\begin{array}{cccc}
c_{13} & 0 & s_{13} & 0\\
0 & 1 & 0 & 0\\
-s_{13} & 0 & c_{13} & 0 \\
0 & 0 & 0 & 1 \end{array} \right)\, ,
R_{23} (\theta_{23})  =  \left (\begin{array}{cccc}
1 & 0 & 0 & 0\\
0 & c_{23} & s_{23} & 0 \\
0 & -s_{23} & c_{23} & 0 \\
0 & 0 & 0 & 1 \end{array} \right)\,\, .
\label{form5}
\eea
}
Therefore $U_{(4 \times 4)}$ takes the form as 
{\small
\bea
U_{(4 \times 4)} &=& \left (\begin{array}{cccc}
c_{14} & 0 & 0 & s_{14} \\
-s_{14}s_{24} & c_{24} & 0 &c_{14}s_{24} \\
-c_{24}s_{14}s_{34} & -s_{24}s_{34} & c_{34} & c_{14}c_{24}s_{34} \\
-c_{24}s_{14}c_{34} & -s_{24}c_{34} & -s_{34} & c_{14}c_{24}c_{34}
\end{array} \right ) \times
\left (\begin{array}{cccc}
{\tilde{U}}_{e1} & {\tilde{U}}_{e2} & {\tilde{U}}_{e3} & 0 \\
{\tilde{U}}_{\mu1} & {\tilde{U}}_{\mu2} & {\tilde{U}}_{\mu3} & 0 \\
{\tilde{U}}_{\tau1} & {\tilde{U}}_{\tau2} & {\tilde{U}}_{\tau3} & 0 \\
0 & 0 & 0 & 1 \end{array} \right ) \nonumber
\eea 
}
{\small
\bea
&=& \left (\begin{array}{cccc}
c_{14}{\tilde{U}}_{e1} & c_{14}{\tilde{U}}_{e2} & c_{14}{\tilde{U}}_{e3} & s_{14}  \\
& & & \\
-s_{14}s_{24}{\tilde{U}}_{e1}+c_{24}{\tilde{U}}_{\mu1} &
-s_{14}s_{24}{\tilde{U}}_{e2}+c_{24}{\tilde{U}}_{\mu2} &
-s_{14}s_{24}{\tilde{U}}_{e3}+c_{24}{\tilde{U}}_{\mu3} & c_{14}s_{24}  \\
&&& \\
\begin{array}{c}
-c_{24}s_{14}s_{34}{\tilde{U}}_{e1}\\
-s{24}s{34}{\tilde{U}}_{\mu1}\\
+c_{34}{\tilde{U}}_{\tau1} \end{array} &
\begin{array}{c}
-c_{24}s_{14}s_{34}{\tilde{U}}_{e2}\\
-s{24}s{34}{\tilde{U}}_{\mu2}\\
+c_{34}{\tilde{U}}_{\tau2} \end{array}  &
\begin{array}{c}
-c_{24}s_{14}s_{34}{\tilde{U}}_{e3}\\
-s{24}s{34}{\tilde{U}}_{\mu3}\\
+c_{34}{\tilde{U}}_{\tau3} \end{array}  &
c_{14}c_{24}s_{34}    \\
&&& \\
\begin{array}{c}
-c_{24}c_{34}s_{14}{\tilde{U}}_{e1}\\
-s_{24}c_{34}{\tilde{U}}_{\mu1}\\
-s_{34}{\tilde{U}}_{\tau1} \end{array}  &
\begin{array}{c}
-c_{24}c_{34}s_{14}{\tilde{U}}_{e2}\\
-s_{24}c_{34}{\tilde{U}}_{\mu2}\\
-s_{34}{\tilde{U}}_{\tau2} \end{array}  &
\begin{array}{c}
-c_{24}c_{34}s_{14}{\tilde{U}}_{e3}\\
-s_{24}c_{34}{\tilde{U}}_{\mu3}\\
-s_{34}{\tilde{U}}_{\tau3} \end{array}  &
c_{14}c_{24}c_{34}  \end{array} \right )\,\,  ,
\label{form6}
\eea
}
where $\tilde{U}_{\alpha i }$ etc. indicate the elements of the flavour mixing 
matrix in 3-flavour scenario, which can be expressed as \cite{amit,mat,mat1}
\bea
\tilde{U} &=& \left (\begin{array}{ccc}
c_{12}c_{13} & s_{12}s_{13} & s_{13} \\
-s_{12}c_{23}-c_{12}s_{23}s_{13} & c_{12}c_{23}-s_{12}s_{23}s_{13} &
s_{23}c_{13} \\
s_{12}s_{23}-c_{12}c_{23}s_{13} & -c_{12}s_{23}-s_{12}c_{23}s_{13} &
c_{23}c_{13}  \end{array} \right )\,\,  .
\label{form7}
\eea
In Eqs. (\ref{form5}-\ref{form7}), $\cos\theta_{ij}=c_{ij}$ and $\sin\theta_{ij}=s_{ij}$  where $\theta_{ij}$ defines the mixing 
angle between $i$th and $j$th neutrinos with mass eigenstates $|\nu_i \rangle$ 
and $|\nu_j \rangle$.

The time evolution equation in the case 
of four neutrino flavours, $|\nu_e \rangle$, $|\nu_\mu \rangle$, 
$|\nu_\tau \rangle$ and $|\nu_s \rangle$ is given by 
\bea
i \displaystyle\frac {d} {dt}\left (\begin{array}{c}
\nu_{e} \\
\nu_{\mu} \\
\nu_{\tau} \\
\nu_{s} \end{array} \right )
&=& H \left (\begin{array}{c}
\nu_{e} \\
\nu_{\mu} \\
\nu_{\tau} \\
\nu_{s} \end{array} \right )\,\, ,
\label{form9}
\eea
where 
\bea
H &=& U_{(4 \times 4)} H_d U^{\dagger}_{(4 \times 4)} \,\, .
\label{form10} 
\eea
In the above, the Hamiltonian in the mass basis is given by
\bea
H_d &=& \left ( \begin{array}{cccc}
E_1 & 0 & 0 & 0 \\
0 & E_2 & 0 & 0 \\
0 & 0 & E_3 & 0 \\
0 & 0 & 0 & E_4 \end{array}
 \right )\,\, ,
\label{form11}
\eea
where $E_i,~i=1-4$ are the energy eigenvalues which can be expressed in terms of the 
momentum $p$ and mass eigenvalues $m_i$, as
\bea
E_i = \sqrt{p_i^{2} +m_i^{2}} \simeq p_i + \displaystyle\frac {m_i^{2}} {2 p_i}
\simeq p + \displaystyle\frac {m_i^{2}} {2 E}\,\, ,
\label{form12}
\eea
with $i = 1,2,3,4$ and $p_i \simeq p,$. With this $H_d$ can be rewritten as 
\bea
H_d &=& \left ( \begin{array}{cccc}
p & 0 & 0 & 0 \\
0 & p & 0 & 0 \\
0 & 0 & p & 0 \\
0 & 0 & 0 & p \end{array}
 \right ) + \displaystyle \frac {1} { 2 E}
\left (\begin{array}{cccc}
m_1^{2} & 0 & 0 & 0 \\
0 & m_2^{2} & 0 & 0 \\
0 & 0 & m_3^{2} & 0 \\
0 & 0 & 0 & m_4^{2} \end{array}
\right)\,\, .
\label{form13}
\eea
In Eq. (\ref{form13}), the matrix ${\rm diag}(p,p,p,p)$ does not contribute to 
the neutrino oscillations as it does not induce any phase differences 
between the neutrinos and hence we do not consider this term further in the 
calculation. Subtracting $m_1^{2}$ from all the diagonal 
elements of the matrix diag$(m_1^{2},m_2^{2},m_3^{2},m_4^{2})$, we have 
\bea
H_d &=& \displaystyle\frac {1} {2 E} {\rm diag} (0,\Delta m_{21}^2,\Delta m_{31}^2,\Delta m_{41}^2)\,\, ,
\label{form14}
\eea
where $\Delta m_{21}^2 = m_2^{2} - m_1^{2}, 
\Delta m_{31}^2 = m_3^{2} - m_1^{2}, \Delta m_{41}^2 = m_4^{2} - m_1^{2}$. 

As discussed in Section 1 the violation of equivalence principle can also 
induce neutrino oscillations due to different gravitational couplings to 
different types 
of neutrinos. As the neutrinos of different types couple differently, the 
gravitational constant ($G$) should be different for different types of neutrinos. 
In addition to the mass induced 
oscillations, the gravity eigenstates ($|\nu_{G i} \rangle$) can also lead to 
the 
neutrino oscillations if gravity eigenstates for neutrinos are not identical to their flavour eigenstates. We explore the mass flavour oscillations in matter 
and gravity 
induced oscillations in a single framework by considering 
$|\nu_\alpha \rangle \ne |\nu_i \rangle \ne |\nu_{G i} \rangle$. It is 
discussed in Section 1 that the 
neutrino energies are red-shifted by an amount 
$E \rightarrow E' = \sqrt{g_{00}} E$ with respect to the vacuum with 
$E' = E(1 - \frac { G M} {R}) = E(1 + \phi)$ where $g_{00} = (1 + 2\phi)$, 
$\phi$ being the gravitational potential, $M$ is the mass of the source of the 
gravitational field and $R$ is the distance over which the gravitational 
field operates. In 4-flavour framework, the gravity eigenstates 
$|\nu_{G i} \rangle (i = 1,2,3,4)$ are connected to the flavour eigenstates 
$|\nu_\alpha \rangle (\alpha = e, \mu, \tau, s)$ through a mixing matrix
$U^{\prime}_{(4\times4)}$ with gravity-flavour mixing angle 
$\theta_{ij}^{G} (i \ne j), i,j = 1,2,3,4$ in the presence of the 
gravitational field. Thus
\bea
|\nu_\alpha \rangle &=& U_{(4 \times 4)}^{'} |\nu_{G_i} \rangle \,\, ,
\label{form15}
\eea
where the gravity-flavour mixing matrix ($U_{(4 \times 4)}^{'}$) can be 
represented as 
{\small
\bea
{U}_{(4 \times 4)}^{'} &=& \left (\begin{array}{cccc}
c'_{14}{\tilde{U'}}_{e1} & c'_{14}{\tilde{U'}}_{e2} & c'_{14}{\tilde{U'}}_{e3} & s'_{14}  \\
& & & \\
-s'_{14}s_{24}{\tilde{U'}}_{e1}+c'_{24}{\tilde{U'}}_{\mu1} &
-s'_{14}s_{24}{\tilde{U'}}_{e2}+c'_{24}{\tilde{U'}}_{\mu2} &
-s'_{14}s_{24}{\tilde{U'}}_{e3}+c'_{24}{\tilde{U'}}_{\mu3} & c'_{14}s_{24}  \\
&&& \\
\begin{array}{c}
-c'_{24}s_{14}s'_{34}{\tilde{U'}}_{e1}\\
-s'_{24}s'_{34}{\tilde{U'}}_{\mu1}\\
+c'_{34}{\tilde{U'}}_{\tau1} \end{array} &
\begin{array}{c}
-c'_{24}s'_{14}s'_{34}{\tilde{U'}}_{e2}\\
-s'_{24}s'_{34}{\tilde{U'}}_{\mu2}\\
+c'_{34}{\tilde{U'}}_{\tau2} \end{array}  &
\begin{array}{c}
-c'_{24}s'_{14}s'_{34}{\tilde{U'}}_{e3}\\
-s'{24}s'{34}{\tilde{U'}}_{\mu3}\\
+c'_{34}{\tilde{U'}}_{\tau3} \end{array}  &
c'_{14}c'_{24}s'_{34}    \\
&&& \\
\begin{array}{c}
-c'_{24}c'_{34}s'_{14}{\tilde{U'}}_{e1}\\
-s'_{24}c'_{34}{\tilde{U'}}_{\mu1}\\
-s'_{34}{\tilde{U'}}_{\tau1} \end{array}  &
\begin{array}{c}
-c'_{24}c'_{34}s'_{14}{\tilde{U'}}_{e2}\\
-s'_{24}c'_{34}{\tilde{U'}}_{\mu2}\\
-s'_{34}{\tilde{U'}}_{\tau2} \end{array}  &
\begin{array}{c}
-c'_{24}c'_{34}s'_{14}{\tilde{U'}}_{e3}\\
-s'_{24}c'_{34}{\tilde{U'}}_{\mu3}\\
-s'_{34}{\tilde{U'}}_{\tau3} \end{array}  &
c'_{14}c'_{24}c'_{34}  \end{array} \right )\,\,\,\,\,\,\,\,\,\,\,\,\,\,\,\, .
\label{form16}
\eea
}
In the above, $\tilde{U}'_{\alpha i}$ ($\alpha=e,\mu,\tau;\, i=1,2,3$) are the elements of the 3-neutrino gravity-flavour mixing matrix $\tilde{U'}$, whose term is similar to Eq.~(6) (but the mixing angles may be different from mass-flavour case). As mentioned earlier, the gravity-flavour mixing angles are denoted by $\theta^G_{ij}$. It may be noted that Eq.~(14) is similar to Eq.~(5) where $\theta_{ij}\rightarrow \theta_{ij}^G$, $U_{(4\times 4)}\rightarrow U'_{(4\times 4)}$, $\tilde{U}\rightarrow \tilde{U}'$. Also note that, Eq.~(13) is similar to Eq.~(2) where $U_{(4\times 4)}\rightarrow U'_{(4\times 4)}$ and $|\nu_i \rangle \rightarrow |\nu _G \rangle.$
The evolution equation in flavour basis due to the presence of the gravitational field (only for gravity-flavour oscillation case) is therefore
written as 
\bea
i \displaystyle\frac {d} {dt} |\nu_{\alpha} \rangle &=& H'
|\nu_{\alpha} \rangle \,\, ,
\label{form17}
\eea
where $H^{\prime}={U}'_{(4 \times 4)} H_{G} {U}'^{\dagger}_{(4 \times 4)}$ and for 
4-flavour scenario $H_{G} = {\rm diag} (E_{G1}, E_{G2}, E_{G3}, E_{G4})$ (diagonal in gravity basis $i \dfrac{d}{dt} |\nu_G\rangle =H_G |\nu_G \rangle$). Substituting $E_{G1}$ from other elements of $H_G$, $H_G$ takes the form
\bea
H_G=\left(
\begin{array}{cccc}
	0&0&0&0\\
	0&\Delta E_{21,G}&0&0\\
	0&0&\Delta E_{31,G}&0\\
	0&0&0&\Delta E_{41,G}
\end{array}
\right).
\eea

If the equivalence principle is indeed violated, all the 
gravitational energy eigenvalues will induce phase differences to
neutrino eigenstates and therefore we have 
$$
H_{G} = {\rm diag} 
\left[ (1 - \phi \alpha_{1})E, (1 - \phi \alpha_{2})E, 
(1 - \phi \alpha_{3})E, (1 - \phi \alpha_{4})E) \right].
$$ 
With $\phi \alpha_{i} = 
\frac {G_{i} M} {R} = \frac {G M} {R} {\alpha_{i}}$. In this case, the phase 
differences can be expressed as 
\be
\Delta {E_{ij, G}} = \frac{GM} {R} \Delta{\alpha_{ij}} E = 
\frac{GM} {R} (\alpha_i - \alpha_j) E = \phi \Delta \alpha_{ij} E = \Delta f_{ij} E\,\, ,
\label{form19}
\ee
where $\Delta{f}_{ij} = \displaystyle\frac {G M} {R} \Delta{\alpha}_{ij} = 
\Delta{\alpha}_{ij} \phi; i,j = 1,2,3,4$. After substracting 
$(1 - \phi \alpha_1)E$ term from all the diagonal elements of $H_G$, we have 
$H_G = {\rm diag} (0,\Delta f_{21}E,\Delta f_{31}E, \Delta f_{41}E)$. It is now well established that 
neutrino oscillations in matter may differ 
significantly from that in vacuum and which was 
first observed by Mikheyev - Smirnov - Wolfenstein \cite{msw,msw1}, and 
known as MSW effect. In the present neutrino oscillations formalism we also include the MSW 
effect. The effective Hamiltonian of the system in flavour basis including both gravity effect and matter effect is given by
\bea
H'' &=& H + H' + V\,\, \nonumber\\
&=& {U}_{(4 \times 4)} H_{d} {U}^{\dagger}_{(4 \times 4)} + 
{U}'_{(4 \times 4)} H_{G} {U}'^{\dagger}_{(4 \times 4)} + V \,\, ,
\label{form20}
\eea
such that, $i\dfrac{d}{dt}|\nu_{\alpha}\rangle = H''|\nu_{\alpha}\rangle$. In the above, the matter potential ($V$) can be written as 
\bea
V = {\rm diag}(V_{CC},0,0,-V_{NC})\,\, ,
\eea
where $V_{CC}$ is the charged current 
potential that appears due to the interactions with the electrons of the medium, 
which are mediated by the $W^{\pm}$ exchange and $V_{NC}$ denotes the neutral 
current potential responsible for the interactions mediated by $Z^{0}$ bosons. 
With $V_{CC}=\sqrt{2} G_{F} N_e$ and $V_{NC}= \frac{G_{F} N_n}{\sqrt{2}}$, the matter potential ($V$) can be
 expressed as 
\bea
V &=& {\rm diag}(\sqrt{2}G_FN_e,0,0,G_FN_n/\sqrt{2})\,\, ,
\label{form21}
\eea
where $G_F$ is the Fermi constant, $N_e$ and $N_n$ are the number densities 
of electron and neutrons respectively inside the matter through which neutrinos
propagate. In our formalism, for the purpose of the 
calculation we assume that the mass mixing angles ($\theta_{ij}$) and gravity 
mixing angles ($\theta_{ij}^G$) with the flavour eigenstates are same, and hence
$U_{(4 \times 4)} = U_{(4 \times 4)}' = U$. The effective Hamiltonian 
according to this assumption takes the form
\bea
H'' &=& U (H_d + H_G) U^{\dagger} + V\,\, \nonumber \\
&=& U({\rm diag}(0, \displaystyle\frac {\Delta{m}_{21}^{2}} {2E}, 
\displaystyle\frac {\Delta{m}_{31}^{2}} {2E}, \displaystyle\frac 
{\Delta{m}_{41}^{2}} {2E}) \\ \nonumber
&&+{\rm diag}(0, \Delta{f}_{21} E, \Delta{f}_{31} E, 
\Delta{f}_{41} E)) U^{\dagger} + V \,\, .
\label{form22} 
\eea
We neglect the terms $\Delta{m}_{21}^{2}$ and 
$\Delta{f}_{21}$ by assuming that the neutrino mass eigenstates 
$|\nu_1 \rangle$, $|\nu_2 \rangle$ as well as gravity eigenstates 
$|\nu_{G1} \rangle, |\nu_{G2} \rangle$ are very close to each other. Thus the 
above Eq. (21) can be written as
\bea
H'' &=& U \, {\rm diag} (0, 0, \displaystyle\frac {\Delta{m}_{31}^{2}} {2E} +
\Delta{f}_{31} E, \displaystyle\frac {\Delta{m}_{41}^{2}} {2E} +
\Delta{f}_{41} E) U^{\dagger} + V\nonumber\\
&=& U \, {\rm diag} (0,0,\displaystyle\frac {\Delta{\mu}_{31}^{2}} {2E}, 
\displaystyle\frac {\Delta{\mu}_{41}^{2}} {2E}) U^{\dagger} + V\,\, .
\label{form23}
\eea
In the above,
\bea
\displaystyle\frac {\Delta{\mu}_{31}^{2}} {2E} &=& \displaystyle\frac 
{\Delta{m}_{31}^{2}} {2E} + \Delta{f}_{31} E \,\, , \nonumber\\
\displaystyle\frac {\Delta{\mu}_{41}^{2}} {2E} &=& \displaystyle\frac 
{\Delta{m}_{41}^{2}} {2E} + \Delta{f}_{41} E \,\, .
\label{form24}
\eea
In Eq. (\ref{form23}), the unitary matrix $U$ is the $4\times 4$ matrix 
similar to that given in Eq. (\ref{form6}) (with different mixing angle parameters). The active neutrino mixing angles 
described in 
Eq. (\ref{form7}) are obtained from the latest bounds given by different neutrino experiments 
\cite{Tanabashi:2018oca}. In this work, we use the best fit
values of the standard three neutrino oscillation parameters which are given 
as \cite{Tanabashi:2018oca}
\bea
\theta_{12}= 33.96^0, \hskip 5mm \theta_{23}=48.3^0, \hskip 5 mm 
\theta_{13}=8.61^0 \nonumber \\
\Delta m_{21}^2=7.53 \times 10^{-5} {\rm eV}^{-2}, \hskip 5mm
\Delta m_{31}^2=2.5\times 10^{-3} {\rm eV}^{-2}\,\ .
\label{neu3}
\eea
For simplicity, we also consider the case of normal hierarchy for 
neutrino eigenstates and CP violating phase $\delta_{\rm CP}=0$. 

Apart from the active neutrino
oscillation parameters mentioned in Eq. (\ref{neu3}), there are three 
active-sterile neutrino mixing angles $\theta_{14},~\theta_{24},~\theta_{34}$.
Several neutrino oscillation experiments
such as MINOS \cite{minos}-\cite{DeRijck:2017ynh}, Daya Bay \cite{Adamson:2016jku}-\cite{daya6}, Bugey \cite{bugey},
T2K \cite{Abe:2019fyx}, IceCube \cite{Aartsen:2017bap} etc. provide stringent limits on these
mixing angles ($\theta_{i4},~i=1-3$) for different values of mass square 
difference $\Delta 
m_{41}^2$. We use the combined limit on mixing angle $\theta_{14}$ obtained 
from the analyses by Daya Bay, MINOS and Bugey-3 \cite{Adamson:2016jku}. However,
in a recent work by Adams {\it et al.}, constraints from cosmological data were
also taken into account \cite{Adams:2020nue} along with the neutrino oscillation results
from other experiments. From their analyses, they conclude that 
Planck data exclude the regions with $\Delta m_{41}^2 \geq 5\times 10^{-2}$ eV$^2$
and for  $\Delta m_{41}^2 \leq 5\times 10^{-2}$ eV$^2$, limits from the combined 
analysis of  Daya Bay, MINOS and Bugey-3 \cite{Adamson:2016jku} become significant (see Fig.~2 and Fig.~4
of Ref.~\cite{Adams:2020nue} for details). Using these constraints, in the present work, we adopt $\theta_{14}=3.6^0$
and present our results for two values of  $\Delta m_{41}^2$ namely, 
$\Delta m_{41}^2=1\times 10^{-3}$ eV$^2$
and $\Delta m_{41}^2=3\times 10^{-3}$ eV$^2$, consistent with the latest experimental findings. 
MINOS and MINOS+ \cite{Adamson:2017uda,DeRijck:2017ynh} also provide limits on the active-sterile mixing angle 
$\theta_{24}$. From their analyses, it is found that for $\Delta m_{41}^2\geq 10^{-2}$ eV$^2$, 
MINOS+ provides strong upper bound on the mixing angle $\theta_{24}$. However,
 it is observed that $\theta_{24}\leq 26.7^0$ with the choice
$\Delta m_{41}^2=1\times 10^{-3}$ eV$^2$ and $\theta_{24}\leq 50.7^0$ when 
$\Delta m_{41}^2=3\times 10^{-3}$ eV$^2$. However, recent analysis for the search of 
sterile neutrino performed by T2K far detector with 295 km baseline length 
\cite{Abe:2019fyx} predicts $\theta_{24}\leq 22.7^0$ for 
$\Delta m_{41}^2=1\times 10^{-3}$ eV$^2$ and for $\Delta m_{41}^2=3\times 10^{-3}$ eV$^2$ the limit on mixing angle is $\theta_{24}\leq 15.3^0$. With the above limit on 
$\theta_{24}$ for $\Delta m_{41}^2=3\times 10^{-3}$ eV$^2$, the limit on $\theta_{34}$
  is found to be $\theta_{34}\leq 53.1^0$ (see Fig.~4 of 
Ref.~ \cite{Abe:2019fyx} for details). Therefore, we observe that for smaller 
values of $\Delta m_{41}^2\sim 10^{-3}$ eV$^2$, although the mixing angle 
$\theta_{14}$ is very much constrained, limits on other mixing angles namely 
$\theta_{24},~\theta_{34}$ are not much 
stringent. In this work, we adopt two different sets of 
active-sterile 
neutrino mixing angles given in Table~\ref{1} which are in agreement with different neutrino 
oscillation experimental results for smaller values of $\Delta m^2_{41}\sim 10^{-3}$ eV$^2$.

It is to be noted that apart from the $\delta_{CP}$, the 3+1 scenario involves two new CP
phases $\delta_{14}$ and $\delta_{24}$. However, as mentioned 
in~\cite{Adamson:2016jku}, experiments like MINOS, Daya Bay and Bugey-3 are 
based on  disappearance measurements which makes them insensitive to CP phases.
Apart from that, IceCube~\cite{Aartsen:2017bap} experiment also carry out 
their measurements of active-sterile mixing angle considering CP phases to be zero.
We use various limits on active-sterile mixing provided by
these experiments that are insensitive to the choice of CP phases.
In the present work we assume a CP conserving scenario and set all CP phases 
to zero.
\begin{table}[]
\centering
\caption{Chosen 4-flavour mixing angle parameter sets for the calculation 
of gravity induced 4-flavour oscillations in matter.}
\vskip 2mm
\begin{tabular}{|c|c|c|c|}
\hline
Set & $\theta_{14}$ & $\theta_{24}$ & $\theta_{34}$ \\ \hline
1 & 3.6$^{\circ}$ & 4.0$^{\circ}$ & 18.48$^{\circ}$ \\ \hline
2 & 2.5$^{\circ}$ & 10.0$^{\circ}$ & 30.0$^{\circ}$ \\ \hline
\end{tabular}
\label{1}\end{table}

With the two sets of mixing angles tabulated in Table 1 and neutrino mass square differences
 mentioned above, we now calculate the modified four neutrino oscillations
 probabilities within matter including the effects of violation of equivalence 
 principle with new VEP parameters $\Delta f_{31}$ and $\Delta f_{41}$. The
 Hamiltonian $H''$ is then diagonalised by a new $4\times 4$ unitary matrix
 $U^m$ whose elements are similar to that of the matrix $U$ (as in Eq.~\ref{form6}) but with new 
 modified mixing angles. 
Therefore, the oscillation probability for a neutrino 
$|\nu_\alpha \rangle$ having flavour $\alpha$ oscillate to a neutrino 
$|\nu_\beta \rangle$ of flavour $\beta$ is given by the expression 
\cite{prob,madhu}
\bea
P_{\alpha \beta} &=& \delta_{\alpha\beta}
- 4\displaystyle\sum_{j>i} U^{m}_{\alpha i} U^{m}_{\beta i} U^{m}_{\alpha j} 
U^{m}_{\beta j}
\sin^2\left (\frac {\pi L} {\lambda_{ij}} \right )\,\, ,
\label{form25}
\eea
where $U^{m}_{\alpha i}$ etc. are the matrix elements of the unitary matrix 
($U^{m}$), which is computationally obtained by diagonalising the effective 
Hamiltonian $H''$ in Eq. (\ref{form23}) and $L$ indicates the 
baseline length. The oscillation length ($\lambda_{ij}$) in the presence of 
both mass and gravity induced oscillations in matter can be expressed as 
%\bea
%%\lambda_{ij} = \displaystyle\frac {2 \pi} {\Delta{\mu}_{ij}^{2}/2E} = 
%\displaystyle\frac {2 \pi} {\left ( \Delta{m}_{ij}^{2}/2E + \Delta{f}_{ij}
%E \right )} \,\, .
%\label{form26}
%
%\eea
\bea
\lambda_{ij} = \displaystyle\frac {2 \pi} {E'_j - E'_i} = 
\displaystyle\frac {2 \pi} {\Delta E'_{ij}} \,\, ,
\label{form26}
\eea
where $E'_i, E'_j(i,j = 1,2,3,4; i \neq j)$ 
are the eigenvalues of the effective 
Hamiltonian $H''$ (Eq. (\ref{form23})). %Now Eq. (\ref{form25}) takes the form 
%\bea
%P_{\nu_\alpha \rightarrow \nu_\beta} &=& \delta_{\alpha\beta}
%- 4\displaystyle\sum_{j>i} U^{m}_{\alpha i} U^{m}_{\beta i} U^{m}_{\alpha j} 
%U^{m}_{\beta j}
%\sin^2\left (\displaystyle\frac {\Delta E_{ij} L} {2} \right )\,\, ,
%\label{form27}
%\eea
%where $S_{ij}^{2} = \sin^2 \left [\left ( \displaystyle\frac {\Delta{m}_{ij}^{2}} {4 E} + \displaystyle\frac {\Delta{f}_{ij} E} {2} \right ) L \right ]$. 
Since the mass eigenstates $|\nu_1 \rangle$ and $|\nu_2 \rangle$ 
can be assumed to be almost degenerate ($\Delta m_{21}^2\sim 10^{-5}$ eV$^2$), 
we have $\Delta m_{31}^2\simeq \Delta m_{32}^2$, $\Delta m_{41}^2\simeq \Delta 
m_{42}^2$. We follow similar convention for neutrino gravity
eigenstates, such that  $|\nu_{G1} \rangle$ and $|\nu_{G2} \rangle$ are also
almost degenerate ($\Delta f_{21}= 0$) and adopt $\Delta{f}_{31} \simeq \Delta{f}_{32}$ and
$\Delta{f}_{41} \simeq \Delta{f}_{42}$. In the following , we explicitly furnish the expressions for 4-neutrino oscillation probabilities for the cases of $\nu_{\mu}\rightarrow \nu_{\mu}$ and $\nu_{e}\rightarrow \nu_{\mu}$ oscillations.
\bea
P_{\mu\mu}^4 &=& 1 - 4\left[|U^{m}_{\mu1}|^2|U^{m}_{\mu2}|^2 \sin^2\left (\displaystyle\frac {\Delta E'_{12} L} {2} \right ) 
+ |U^{m}_{\mu1}|^2|U^{m}_{\mu3}|^2 
\sin^2\left (\displaystyle\frac {\Delta E'_{13} L} {2} \right ) + \right. \nonumber\\
&&\left. |U^{m}_{\mu1}|^2|U^{m}_{\mu4}|^2 \sin^2\left 
(\displaystyle\frac {\Delta E'_{14} L} {2} \right ) + |U^{m}_{\mu2}|^2
|U^{m}_{\mu3}|^2 
\sin^2\left (\displaystyle\frac {\Delta E'_{23} L} {2} \right ) + \right. \nonumber\\
&&\left.|U^{m}_{\mu2}|^2|U^{m}_{\mu4}|^2 \sin^2\left (\displaystyle\frac 
{\Delta E'_{24} L} {2} \right ) + |U^{m}_{\mu3}|^2|U^{m}_{\mu4}|^2 
\sin^2\left (\displaystyle\frac {\Delta E'_{34} L} {2} \right )\right] \nonumber\\
P_{e\mu}^4 &=& 4\left[|U^{m}_{e1}||U^{m}_{\mu 1}||U^{m}_{e2}||U^{m}_{\mu 2}| 
\sin^2\left (\displaystyle\frac {\Delta E'_{12} L} {2} \right ) +
|U^{m}_{e1}||U^{m}_{\mu 1}||U^{m}_{e3}||U^{m}_{\mu 3}| \sin^2\left 
(\displaystyle\frac {\Delta E'_{13} L} {2} \right )+ \right.\nonumber\\
&&\left. |U^{m}_{e1}||U^{m}_{\mu 1}||U^{m}_{e4}||U^{m}_{\mu 4}|) \sin^2\left 
(\displaystyle\frac {\Delta E'_{14} L} {2} \right ) + 
|U^{m}_{e2}||U^{m}_{\mu 2}||U^{m}_{e3}||U^{m}_{\mu 3}| \sin^2\left 
(\displaystyle\frac {\Delta E'_{23} L} {2} \right )+ \right. \nonumber\\
&&\left. |U^{m}_{e2}||U^{m}_{\mu 2}||U^{m}_{e4}||U^{m}_{\mu 4}| \sin^2\left 
(\displaystyle\frac {\Delta E'_{24} L} {2} \right ) + |U^{m}_{e3}||U^{m}_{\mu 3}||U^{m}_{e4}||U^{m}_{\mu 4}| 
\sin^2\left (\displaystyle\frac {\Delta E'_{34} L} {2} \right )\right]\nonumber\\
\label{form28}
\eea
The expressions for other probabilities can similarly be written. It is to be noted that, earlier studies of violation of equivalence principle 
with IceCube neutrino data by Esmaili {\it et. al} set a stringent bound on $\Delta f_{31}\leq 7\times 
10^{-27}$ \cite{Esmaili:2014ota}. Therefore, in the present work, we consider $\Delta f_{31}= 5\times 
10^{-27}$ which is in agreement with previous analysis. However, since the value 
of $\Delta f_{31}$ is negligibly small, we do not expect any significant VEP effect due to $\Delta f_{31}$.
On the other hand there exists no bound on the VEP associated with the sterile neutrino $\Delta f_{41}$. In the 
next section, we investigate how VEP induced by the sterile neutrino affects the 
four flavour neutrino oscillations in matter.

\section{Calculations and Results}

\subsection{Gravity-induced neutrino oscillations in matter}

In this section, we study modification of neutrino oscillation 
probabilities in matter in presence of possible violation of equivalence 
principle. For this purpose, we consider a 4-flavour neutrino scenario, where the usual three families of 
active neutrinos ($\nu_e, \nu_\mu, \nu_\tau$) are extended by an extra setrile 
neutrino ($\nu_s$). In order to estimate the effect of the gravity induced 
oscillations, we consider baseline neutrino oscillations, where the neutrinos 
are produced from a neutrino 
factory or an accelerator and propagate from the source to a far terrestrial 
away detector through the Earth matter. 

The probabilities of oscillations from one flavour to the other in the present 
framework are therefore an important component for the estimation of the 
neutrino flux at the end detector. Therefore we calculate the oscillation 
probabilities from one neutrino flavour to the other for the present 4-neutrino scenario where both 
the mass induced and gravity induced oscillations are 
considered. The probabilities are computed using the Eqs. (\ref{form20}) - (\ref{form28}). We 
demonstrate in this section how the nature of the probabilities are varied 
by the combined effect of the gravity induced factors as well as the 
mass-flavour oscillations in matter. For the present calculations we have 
chosen a demonstrative 
baseline length of 7000 Km and the mean Earth matter density to be 4.15 gm/cc.
From Eq. (\ref{form25}) and Eq. (\ref{form28}) it is clear that the oscillatory part of the 
probability equations are controlled by the phase factor $\Delta E_{ij}' L/2$, 
where $\Delta E_{ij}'$ are the difference of the eigenvalues $E_i'$ and 
$E_j'$ of the 
eigenstates designated by $i$ and $j$ respectively. In this work, the 
eigenvalues $E_i', E_j'$ etc. are computationally obtained by explicitly 
diagonalising the Hamiltonian $H''$  (Eq. (\ref{form23})) that includes both 
the mass induced effects, matter effects as also the gravity induced effects.

\begin{figure}[h!]
	\centering
	\subfigure[Variation of $P_{e\mu}^4$ for $\Delta m_{41}^2 = 1\times 10^{-3}$  eV$^2$]{
		\includegraphics[width= 0.45\linewidth,angle=0]{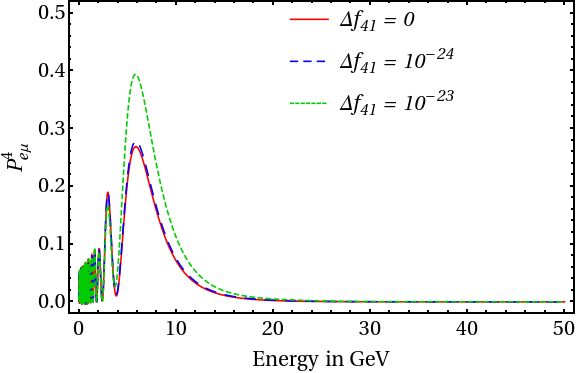}}
	\subfigure [Variation of $P_{\mu\mu}^4$ for $\Delta m_{41}^2 = 1\times 10^{-3}$  eV$^2$]{
		\includegraphics[width= 0.45\linewidth,angle=0]{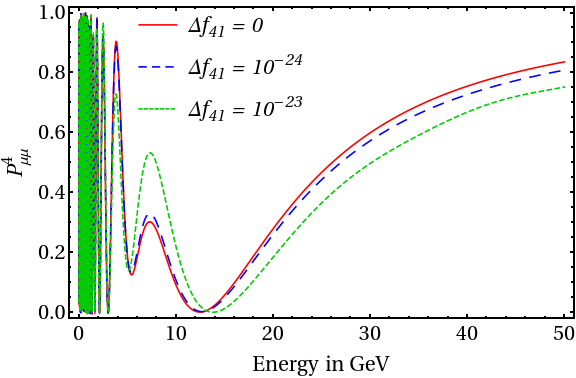}}
	\subfigure[Variation of $P_{e\mu}^4$ for $\Delta m_{41}^2 = 3\times 10^{-3}$  eV$^2$]{
		\includegraphics[width= 0.45\linewidth,angle=0]{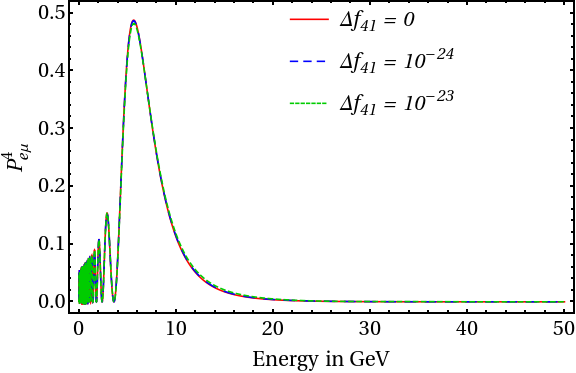}}
	\subfigure[Variation of $P_{\mu\mu}^4$ for $\Delta m_{41}^2 = 3\times 10^{-3}$  eV$^2$]{
		\includegraphics[width= 0.45\linewidth,angle=0]{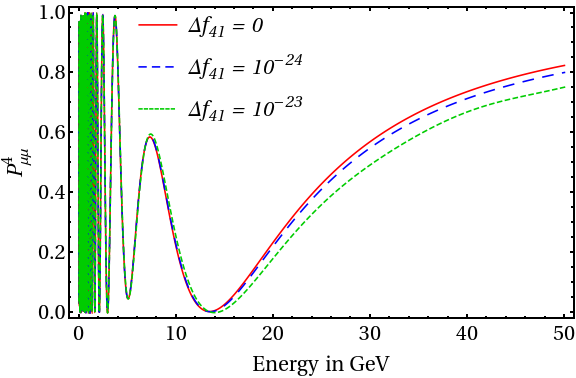}}
	%\subfigure[Probability $P_{\tau \tau}$ as a function of energy at L=7000Km.]{
	%\includegraphics[scale =0.25,angle=-90]{ttm.eps}}
	\caption{Neutrino oscillation probabilities in matter for a fixed value of $\Delta f_{31} = 5 \times 10^{-27}$ and for Set-1 with baseline length $L$ = 7000 km.}
	\label{fig1}
\end{figure}

\begin{figure}[h!]
	\centering
	\subfigure[Variation of $P_{e\mu}^4$ for $\Delta m_{41}^2 = 1\times 10^{-3}$  eV$^2$]{
		\includegraphics[width= 0.45\linewidth,angle=0]{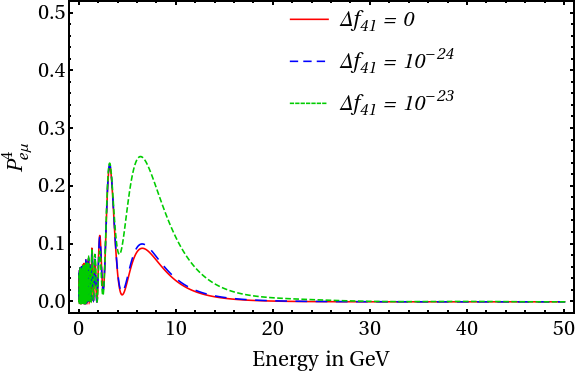}}
	\subfigure [Variation of $P_{\mu\mu}^4$ for $\Delta m_{41}^2 = 1\times 10^{-3}$  eV$^2$]{
		\includegraphics[width= 0.45\linewidth,angle=0]{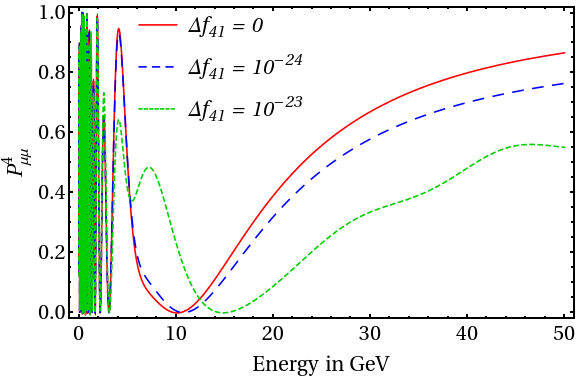}}
	\subfigure[Variation of $P_{e\mu}^4$ for $\Delta m_{41}^2 = 3\times 10^{-3}$  eV$^2$]{
		\includegraphics[width= 0.45\linewidth,angle=0]{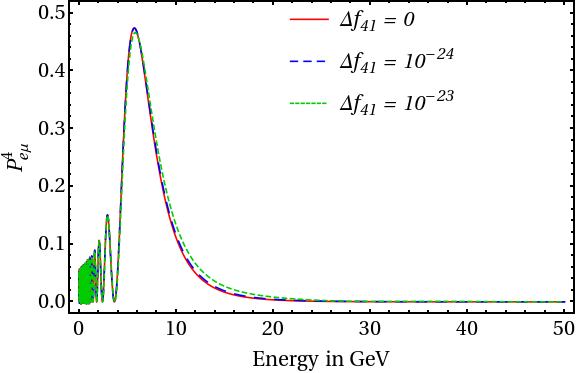}}
	\subfigure[Variation of $P_{\mu\mu}^4$ for $\Delta m_{41}^2 = 3\times 10^{-3}$  eV$^2$]{
		\includegraphics[width= 0.45\linewidth,angle=0]{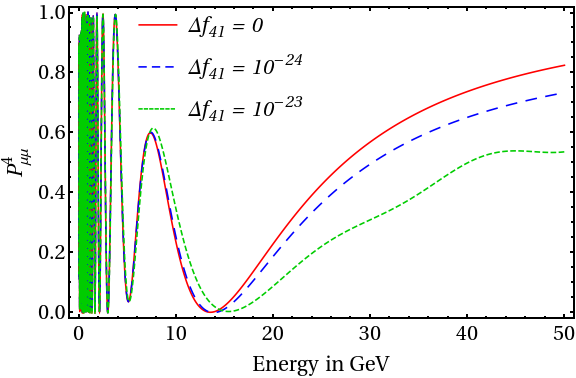}}
	%\subfigure[Probability $P_{\tau \tau}$ as a function of energy at L=7000Km.]{
	%\includegraphics[scale =0.25,angle=-90]{ttm.eps}}
	\caption{Neutrino oscillation probabilities in matter for a fixed value of $\Delta f_{31} = 5 \times 10^{-27}$ and for Set-2 with baseline length $L$ = 7000 km.}
	\label{fig2}
\end{figure}

In Figs.~\ref{fig1}-\ref{fig2}, we show how the oscillation probabilities vary for the variation 
of gravity induced effects (designated by $\Delta f_{ij}$), when the mass square 
difference $\Delta m_{ij}^2$ are kept fixed. Since we are considering here a 
4-flavour scenario, these probability plots also demonstrate the 4-flavour 
oscillations considering both the mass flavour effects in matter and the 
gravity induced effects. 

Figs.~\ref{fig1}-\ref{fig2} show the variations of the probability $P_{e\mu}^4$ and $P_{\mu\mu}^4$ with different neutrino energies for different fixed chosen values of $\Delta f_{41}$. Three values of $\Delta f_{41}$ are chosen namely $\Delta f_{41}=0$, $10^{-24}$, $10^{-23}$. Needless to mention that $\Delta f_{41}=0$. Fig.~\ref{fig1} shows the results obtained using the parameter for Set-1 whereas, in Fig.~\ref{fig2} the probability plots for Set-2 are shown. In both Fig.~\ref{fig1} and Fig.~\ref{fig2}, the upper panels correspond to $\Delta m_{41}^2=1\times 10^{-3}$ eV$^2$, while for the lower panels $\Delta m_{41}^2=3\times 10^{-3}$ eV$^2$. All the computations are made by adopting a fixed value of $\Delta f_{31}=5\times 10^{-27}$. Similar plots for other probabilities (e.g. $P_{ee}^4$, $P_{e\tau}^4$, $P_{\mu\tau}^4$ etc.) can also be computed.

It is to be noted from Figs.~\ref{fig1}-\ref{fig2}, that the VEP effect is more prominent when $\Delta f_{41}=10^{-23}$ and $\Delta m_{41}^2=1\times 10^{-3}$ eV$^2$. It can also be noted that, if $\Delta m_{41}^2=1\times10^{-3}$ eV$^2$, the variations for $P_{e\mu}^4$ with $\Delta f_{41}=10^{-23}$ can be more than 50\% for other chosen values of $\Delta f_{41}$ when Set-2 is adopted. Similar trends are also observed for $P_{\mu\mu}^4$. In the later case, the oscillation is more prominent for Set-2 than Set-1 when $\Delta f_{41}=10^{-23}$ and $\Delta m_{41}^2=1\times 10^{-3}$ eV$^2$ are chosen. It is to be mentioned that all plots of Figs.~\ref{fig1}, \ref{fig2}, the base length $L$ is chosen to be 7000 km. Thus it is demonstrated from Figs.~\ref{fig1}-\ref{fig2} that the probabilities are most affected by the gravity induced effects when $\Delta f_{41} = 10^{-23}$.

\subsection{Effect of Gravity Induced Oscillation on a Long Baseline Neutrino Experiment}

In this section we pursue the effects of gravity induced oscillations on 
neutrino induced muon yields in long baseline (LBL) experiments.
In a long baseline neutrino experiment, pions are initially produced in neutrino 
factories
by directing a proton beam incident on a target. Pions decay into muons which  
suffer further decay in a muon storage ring producing neutrinos. Neutrinos are generated from 3-body
decay of muons as 
\footnote{Instead of the muon storage ring $\beta$ beams from $\beta$-decay of nucleons can also be treated 
as the source of neutrinos. In this case one should consider the channels $P_{\nu_e \rightarrow \nu_x}$, where
$x=e,~\mu,~\tau$.}
\bea
\mu^{-} \rightarrow e^{-} + \bar{\nu}_e +\nu_{\mu} \, ,\\ \nonumber
\mu^{+} \rightarrow e^{+} + \nu_e +\bar{\nu}_{\mu} \,\ .
\eea
Neutrinos produced in neutrino factory are then directed towards a neutrino detector far away
from the source of the neutrinos and traverse through Earth matter to reach the detector. The muon 
neutrinos (${\nu}_{\mu},~\bar{\nu}_{\mu} $) generated in neutrino factory will suffer oscillations
due to its passage through the Earth matter along the baseline. The 
$\nu_{\mu} (\bar{\nu}_{\mu})$
will produce $\mu^-~(\mu^+)$ at the detector by charged current (CC)
 interaction with the detector material. If it is pure $\mu^-$ at the source 
 then only $\nu_{\mu}$ beam will propagate along the baseline and $\mu^-$ will 
 be produced at the detector end which the latter would detect. Those muons 
 are called right sign muon. Needless to mention that $\nu_\mu$ flux at the source will
 suffer depletion due to the oscillation and same will happen to the muon 
yield. 
However if the detector detects a $\mu^+$ instead, then it must be 
that $\bar{\nu}_\mu$ reaches the detector and $\bar{\nu}_\mu$ can only be 
created in the beam (produced by the decay of $\mu^-$) through the oscillation 
$\bar{\nu}_e \rightarrow \bar{\nu}_\mu$ during the passage of  $\bar{\nu}_e$
through the baseline. These events are termed as wrong sign muon events.
The situation is just reversed if $\bar{\nu}_\mu$ beam is produced at the 
storage 
ring from the decay of $\mu^+$. But the right sign and the wrong sign muon events can be
distinguished by a iron calorimeter detector (such as the one considered for 
the present work) when the ICAL detector is magnetized.

For this purpose, we need to compute the probabilities $P^4_{\bar{e}\bar{\mu}}$, where $P^4_{\bar{e}\bar{\mu}}$ is the $\nu_{\bar{e}}$ to $\nu_{\bar{\mu}}$ oscillation probability. It is to be noted that, for the case of anti-neutrinos the matter potential $V$ (see earlier) changes sign. Therefore the computation of $P_{\bar{e}\bar{\mu}}^4$ is done by considering $V$ to $-V$ in the relevant equations described in sect.~2.

As mentioned above, we first consider the neutrino (anti-neutrino) flux  in the
neutrino factory which is expressed as \cite{geer,donini}

\bea \frac{ d^2 \Phi_{\nu_{\mu},
\bar\nu_{\mu}} }{ dy dA}  =  \frac{ 4 n_\mu }{ \pi L^2 m_\mu^6 }
\,\,  E_\mu^4 y^2 \, (1 - \beta) \,\,  \left
[ 3 m_\mu^2 - 4  E_\mu^2 y \, (1 - \beta) \right ]  
\label{eq29}
\eea
and similarly $\nu_e$ ($\bar {\nu}_e$) flux is given by
\bea \frac{ d^2 \Phi_{\nu_e,
\bar\nu_e} }{ dy dA}  =  \frac{ 24 n_\mu }{ \pi L^2 m_\mu^6 }
\,\,  E_\mu^4 y^2 \, (1 - \beta) \,\,  \left
[  m_\mu^2 - 2  E_\mu^2 y \, (1 - \beta) \right ].  
\label{eq30}
\eea
where different terms are given as follows
\begin{itemize}
  \item $E_{\mu} :$ muon energy
  \item $n_{\mu} :$  number of injected  muons
  \item $L :$ distance between neutrino factory and the end detector (baseline length)
  \item $y = \frac{E_{\nu}}{E_{\mu}}$ where ${E_{\nu}}$ is energy of neutrino
  \item $\beta$ is the boost factor
\end{itemize}
It is to be noted that the expressions for neutrino fluxes in 
Eqs.~(\ref{eq29}-\ref{eq30}) are derived
under the following approximations; i) neutrinos are not polarised and 
ii) the angle between direction
of neutrino beam towards the detector and the beam axis is assumed to be zero. 
For the computations of neutrino flux using Eqs. (28,29), we consider 
$\sim 10^{21}$ protons on target per year and muon injection energy of
50 GeV. In Fig.~\ref{flxfgr} we show the flux for $\nu_\mu$ and $\bar{\nu_e}$. 

%For the present study we consider average Earth density $\rho=4.15$ gm/cc. 

\begin{figure}[h!]
\centering
\subfigure[]{
\includegraphics[width=7 cm,scale= 0.5,angle=0]{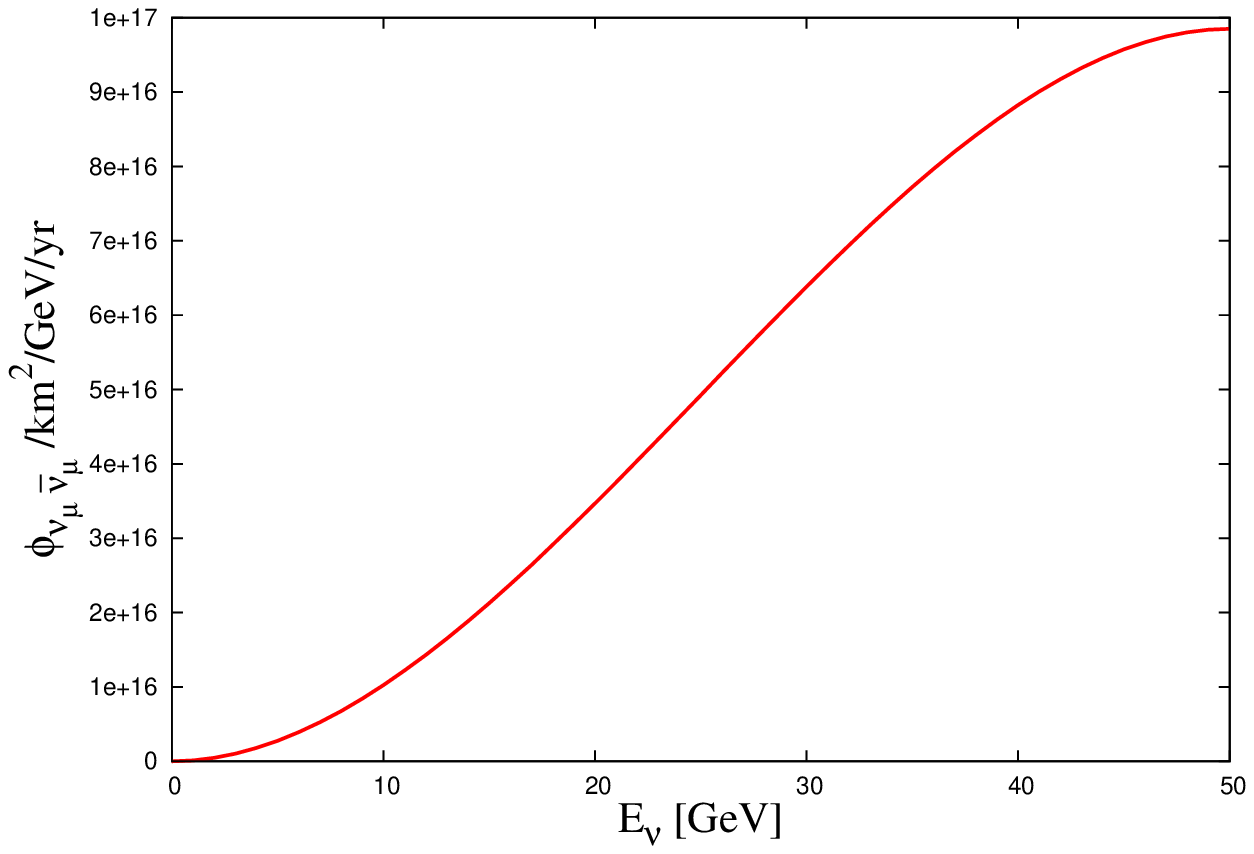}}
\subfigure[]{
\includegraphics[width=7 cm,scale= 0.5,angle=0]{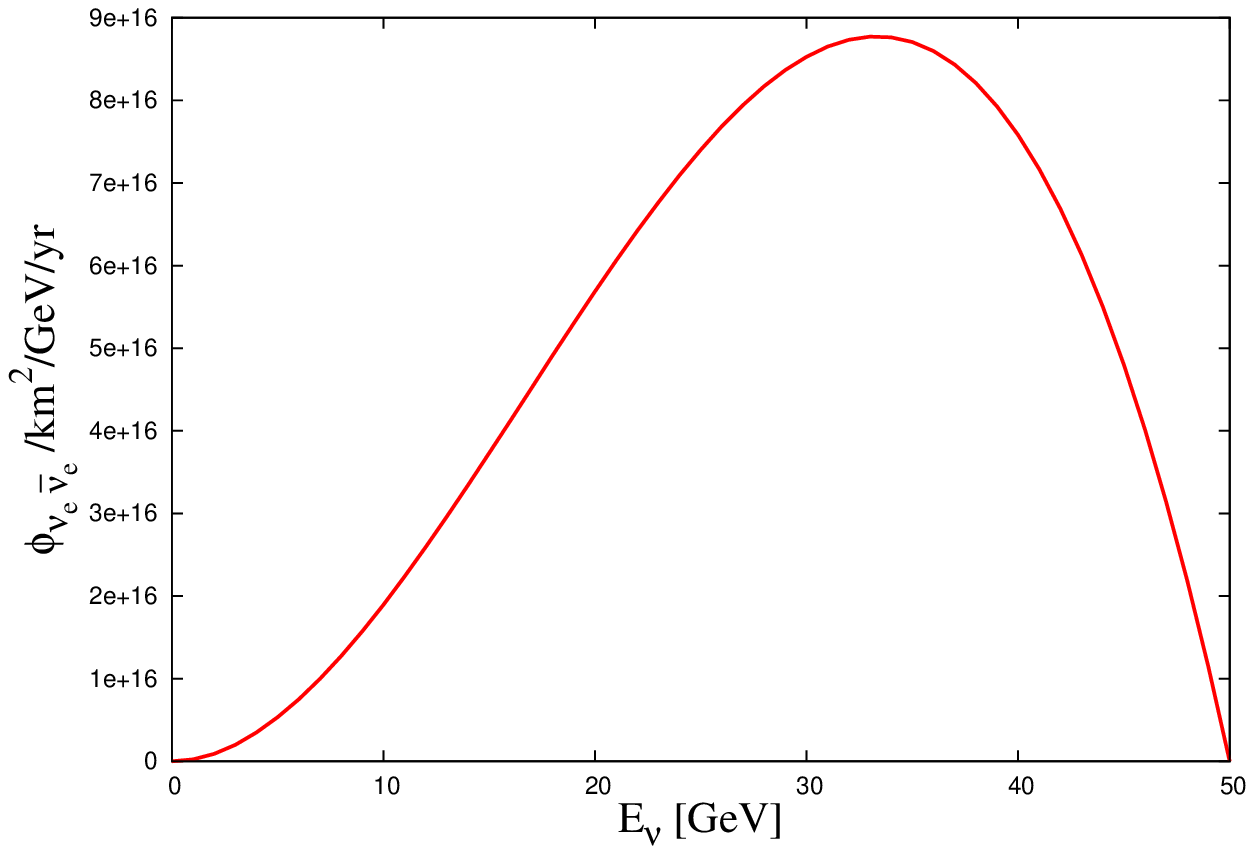}}
\caption{Flux of (a) $\nu_\mu$  and (b) $\bar{\nu_e}$ for muon decay at 
muon storage ring with muon injection energy of 50 GeV. See text for details.}
\label{flxfgr}
\end{figure}

As mentioned, in this work we consider the detector to be a magnetized iron calorimeter (ICAL) 
detector. Such a detector of 50 kTon mass has been suggested for the proposed 
India-based Neutrino Observatory (INO) \cite{Kumar:2017sdq}. The proposed ICAL
detector at INO \cite{Kumar:2017sdq} is basically a stack of 151 iron plates of thickness 5.6 cm and
each plates are separated by a gap of 4 cm containing a total of 
50 kTon of detector iron. Here we consider a baseline length of 7359 km which 
is roughly the distance 
between CERN and proposed INO site. Beam of $\nu_{\mu}~(\bar{\nu}_{\mu} )$
from a neutrino factory
after reaching such a magnetized ICAL detector will undergo charged current interactions and produce
$\mu^-~(\mu^+)$ which form muon tracks while passing through different layers 
of the detector of different curvature due to magnetic field. Observing the 
direction and curvature of the muon tracks one can
distinguish the right sign and wrong sign muons inside detector. 
As mentioned, the flux of neutrino (anti-neutrino) beam will undergo
gravity induced and mass induced four flavour oscillations in matter before 
reaching the detector. Thus the neutrino (anti-neutrino) flux at the detector 
will be rescaled by the corresponding
probabilities.  For $\nu_\mu$ and $\bar{\nu}_{e}$ beam (produced from the 
decay $\mu^-$ at the storage ring) if $\mu^+$ is registered in ICAL then this is
referred to as appearance channel since it originates due the oscillation 
$\bar{\nu}_e \rightarrow \bar{\nu}_{\mu}$ while for the same beam $\mu^-$ 
track is considered as disappearance channel as $\nu_\mu$ disappears via the 
oscillation $\nu_{\mu}\rightarrow \nu_{x}, x\neq \mu$. In this section, we present the
expected yield of right sign muon ($\mu^-$) and wrong sign muon ($\mu^+$)
at ICAL detector in presence of gravity induced neutrino oscillations in four flavour scenario.

\begin{figure}[h!]
\centering
\includegraphics[width=7 cm,scale= 0.5,angle=0]{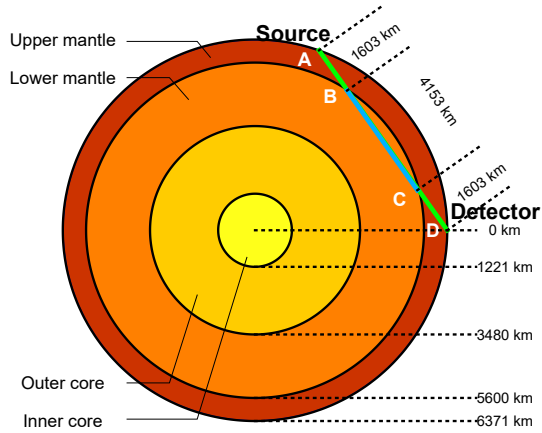}
\caption{Different layers of earth and projected travel path of neutrinos for
long basline neutrino detector placed at $7359$ km from source.}
\label{earth}
\end{figure}

Different neutrino oscillation probabilities calculated in previous 
section are presented assuming average Earth density $\rho=4.15$ gm/cc. 
However for the present scenario, where the baseline length $\sim 7359$ km, neutrinos travel through
Earth crust and mantle. Width of Earth crust is very small $\sim$ 10-15 km 
compared to the long baseline oscillation length 7359 km.
Therefore, oscillation effects due to crust can be ignored safely and we 
consider neutrino oscillation through upper and lower mantle only, as demonstrated
in Fig.~\ref{earth}. Density of upper (lower) mantle of Earth is $\rho_{up}=3.9$ gm/cc
($\rho_{low}=4.5$ gm/cc) \cite{Dziewonski:1981xy}. From Fig.~\ref{earth}, we observe that for the chosen 
long baseline length, neutrinos traverse through upper mantle initially for 
first 1603 kms and then enters lower mantle where it travels a distance of 4153 kms and finally 
enters again into the upper mantle through which they
travel another 1603 kms to reach the detector. We use the above two layer
formalism to calculate  neutrino oscillation probabilities within Earth
for the calculation of right sign and wrong sign muon events at the detector.
For this purpose we need to evaluate new probability amplitudes for neutrino
oscillation within two layers of mantle. For example, the probability amplitude for
the channel $\nu_l \rightarrow \nu_{l'}$ with the two layers of mantle considered 
is expressed as
\bea 
A_{ll'}=\sum_{k,k',k'',\alpha,\beta} A_{lk}A_{kk}^{up}(d)A_{k\alpha}A_{\alpha 
k'}A_{k'k'}^{low}(D)A_{k'\beta}A_{\beta k''}A_{k''k''}^{up}(d)A_{k''l'}
\label{amplitude}
\eea 
where $l,l',\alpha,\beta=e,\mu,\tau,s$; $k,k',k''=1,2,3,4$
and superscripts up (low) correspond to upper and lower mantle with density
$\rho_{up}$ ($\rho_{low}$). 
Eq.~\ref{amplitude} can be explained as follows. The matter effect on neutrinos
as they pass through the matter is related to neutrino interaction 
with matter. The coherent neutrino weak interaction scattering with 
matter proceeds via weak interaction eigen states or flavour eigen
states of neutrinos and depends on the particle density (and hence
matter density) inside the medium through which the neutrino is 
propagating (as discussed earlier in Section 2). But it is the 
mass eigenstates of neutrino in matter (within that medium) 
which propagate through a distance in that medium. 
The neutrino in this case, is produced in a 
particular flavour eigen state in a source and  
enters the  earth matter in the upper mantle with a certain density, through 
which it will first propagate. The possible mass eigenstates (in matter 
of the upper mantle) in the initial flavour eigenstate (due to neutrino 
mixing) is of relevance here. These mass 
eigenstates initially propagate a distance of $d=1603$ km inside 
the upper mantle  
as $\exp(-iEd)$ ($E$ represents the energy eigen value of the neutrino; the 
propagation Hamiltonian is diagonal in mass basis) till they reach 
the boundary of upper and lower mantle
(Fig.~\ref{earth}). 
As the lower mantle has a different matter density,  
the neutrino needs to be converted from its mass eigen state (in 
upper mantle matter) to possible flavour eigenstates which then enter 
the lower mantle. But since neutrino will now propagate a distance 
of $D=4153$ kms inside the lower mantle with matter density different 
from that of the upper mantle, 
these are the mass eigen state(s) in lower mantle matter 
(and not the flavour eigenstate(s)) that are relevant since these mass 
eigen states will now propagate as $\exp(-iED)$ till they reach the 
boundary of lower mantle and upper mantle (Fig.~\ref{earth}). As they 
propagate from lower mantle to upper mantle another change of matter 
density will occur. Following the similar 
procedure, finally, the neutrino mass eigen states in the upper mantle 
will reach the detector on earth after traversing a distance of $d=1603$ kms
(see Fig.~\ref{earth}). But again, since the neutrinos will undergo weak interaction
with detector material inside the detector (and a neutrino of a particular 
flavour will be detected), the relevant flavour eigen state is to be 
obtained from the mass eigenstates that reach the detector. 
In Eq.~(\ref{amplitude}), $A_{lk}=\langle\nu_l|\nu_k\rangle=U_{lk}$ is the element
of neutrino mixing matrix before oscillation within matter 
whereas $A_{k\alpha}\rightarrow U_{k \alpha} (\theta_{up}),~
A_{\alpha k'}\rightarrow U_{\alpha k'} (\theta_{low}),~
A_{k'\beta}\rightarrow U_{k' \beta} (\theta_{low}),~
A_{\beta k''}\rightarrow U_{\beta k''} (\theta_{up})$
corresponds to oscillation within matter
and $A_{kk}^{up}=\langle\nu_k|\nu_k^{up}(d)\rangle=e^{-iEd}$
with $E$ being a function of $\Delta m_{ij}^2,~\Delta f_{ij}, V$.
%In the above expression of probability amplitude, $d$ and $D$ refers to the oscillation length within upper and lower mantle ofEarth. 
Finally the oscillation probability $\nu_l \rightarrow \nu_{l'}$ ($l,l'$ are two
different neutrino flavours) between two different
neutrino flavours can be calculated from the amplitude $P_{ll'}=|A_{ll'}|^2$.

%\begin{figure}[h!]
%\centering
%\subfigure[]{
%\includegraphics[width=7 cm,scale= 0.5,angle=0]{S1_001e_mu_new.eps}}
%\subfigure[]{
%\includegraphics[width=7 cm,scale= 0.5,angle=0]{S1_001mu_mu_new.eps}}
%\caption{Neutrino oscillation probabilities within matter for varying density 
%with two layer approximation and baseline length $L=7359$ km.
%See text for details.}
%\label{newosc}
%\end{figure}

\begin{table}[]
\centering
\caption{The right sign $\mu$ yield and the wrong sign $\mu$ yield in the 
presence of gravity induced 4-flavour oscillations in matter for Set-1 and 
for the fixed values of $\Delta f_{31} = 5 \times 10^{-27}$. The muon injection energy is fixed
at 50 GeV. See text for details.}
%\vskip 2mm
\begin{tabular}{|c|c|c|c|}
\hline
$\Delta m_{41}^2$ in eV$^2$ & $\Delta f_{41}$ & Right sign $\mu$ & Wrong sign $\mu$ \\ 
\hline
\multirow{3}{*}{$1\times 10^{-3}$} & 0  & 3115192 &  4182 \\ 
& 10$^{-24}$ & 2973398 & 4176\\ 
& 10$^{-23}$ & 2662703 & 4543 \\ \hline
\multirow{3}{*}{$3\times 10^{-3}$} & 0  & 3006510 &  4138 \\ 
& 10$^{-24}$ & 2896293 & 3733 \\ 
& 10$^{-23}$ & 2654969 & 5035 \\ \hline 
\end{tabular}
\label{2}
\end{table}

\begin{table}[]
\centering
\caption{The right sign $\mu$ yield and the wrong sign $\mu$ yield in the 
presence of gravity induced 4-flavour oscillations in matter for Set-2 and for the fixed 
values of $\Delta f_{31} = 5 \times 10^{-27}$ with injected muon energy fixed 
at 50 GeV. 
See text for details.}
%\vskip 2mm
\begin{tabular}{|c|c|c|c|}
\hline
$\Delta m_{41}^2$ in eV$^2$ & $\Delta f_{41}$ & Right sign $\mu$ & Wrong sign $\mu$ \\ 
\hline
\multirow{3}{*}{$1\times 10^{-3}$} & 0  & 3386441 &  4501 \\ 
& 10$^{-24}$ & 2891733 &  4500 \\ 
& 10$^{-23}$ & 1832336 &  6139 \\ \hline
\multirow{3}{*}{$3\times 10^{-3}$} & 0  & 3006875 & 4341 \\ 
& 10$^{-24}$ & 2605874 & 3731 \\ 
& 10$^{-23}$ & 1805902 & 6521 \\ \hline 
\end{tabular}
\label{3}
\end{table}

\begin{figure}[h!]
	\centering
	\subfigure[Set-1, $\Delta m_{41}=1\times 10^{-3}$ eV$^2$]{
		\includegraphics[width= 0.45\linewidth,angle=0]{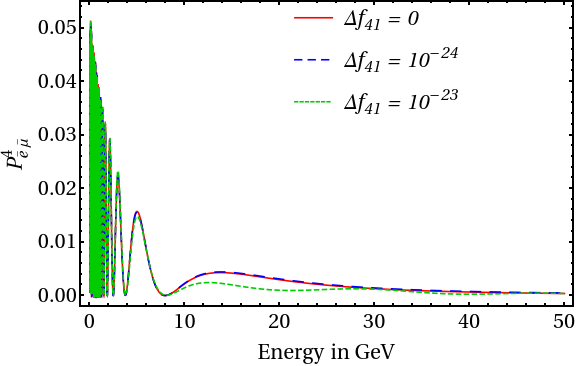}}
	\subfigure [Set-1, $\Delta m_{41}=3\times 10^{-3}$ eV$^2$]{
		\includegraphics[width= 0.45\linewidth,angle=0]{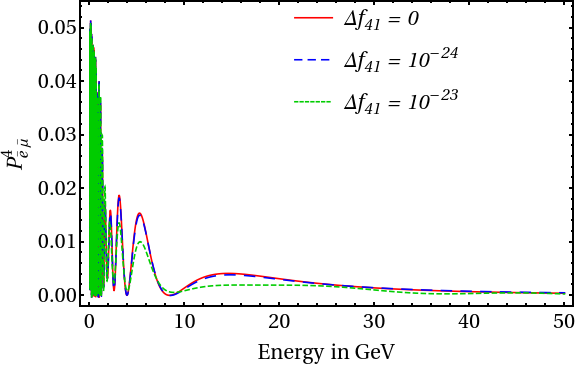}}
	\subfigure[Set-2, $\Delta m_{41}=1\times 10^{-3}$ eV$^2$]{
		\includegraphics[width= 0.45\linewidth,angle=0]{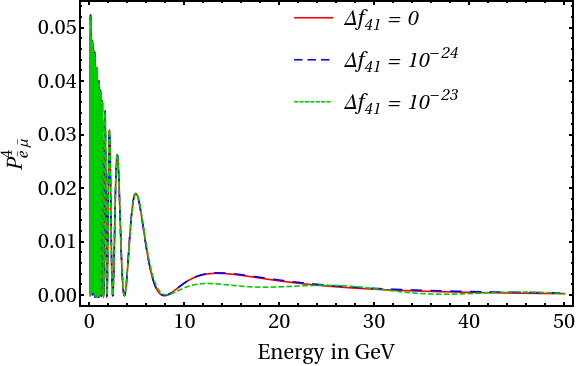}}
	\subfigure[Set-2, $\Delta m_{41}=3\times 10^{-3}$ eV$^2$]{
		\includegraphics[width= 0.45\linewidth,angle=0]{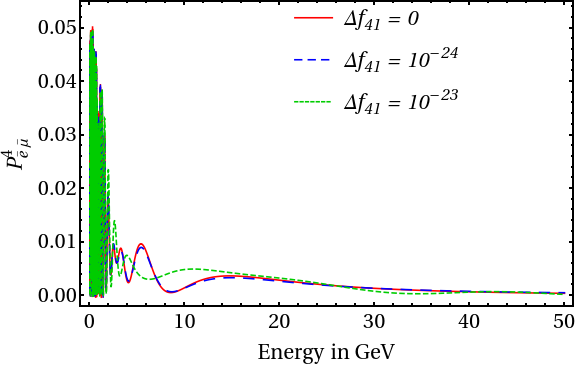}}
	\caption{Neutrino oscillation probabilities $\bar{e}\rightarrow \bar{\mu}$ in matter for baseline length $L$ = 7000 km.}
	\label{figbar}
\end{figure}

The computed probabilities for $P_{\bar{e}\bar{\mu}}^4$ are plotted in Fig.~\ref{figbar} for three chosen values of $\Delta f_{41}$ namely  $\Delta f_{41}=0$, $10^{-24}$ and $10^{-23}$. As in the case of Figs.~\ref{fig1}, \ref{fig2} in Fig.~\ref{figbar} also the upper panel is for Set-1 and the lower panel is for Set-2. Here in the left panel, the computations with $\Delta m_{41}^2=1\times 10^{-3}$ eV$^2$ are shown, while for the right panel $\Delta m_{41}^2=3\times 10^{-3}$ eV$^2$ is adopted. For all the cases, the value of $\Delta f_{31}$ is kept fixed at $\Delta f_{31}=5\times 10^{-27}$. Similar to what is observed in Figs.~\ref{fig1}, \ref{fig2} here also the VEP effect is more prominent when $\Delta f_{41}=10^{-23}$. But contrary to Figs.~\ref{fig1}, \ref{fig2} in Fig.~\ref{figbar}, the VEP effect is prominent when $\Delta m_{41}^2=3\times 10^{-3}$ eV$^2$ is chosen and no appreciable VEP effects are observed when $\Delta m_{41}^2=1 \times 10^{-3}$ eV$^2$. Note that, in Figs.~\ref{fig1}, \ref{fig2} no appreciable VEP effects have been observed for $P_{e\mu}^4$ when $\Delta m_{41}^2=3\times 10^{-3}$ eV$^2$ is chosen, but prominent VEP effects are observed when $\Delta m_{41}^2=1 \times 10^{-3}$ eV$^2$.

In Table~\ref{2}, we furnish the estimated right sign and wrong sign 
muon yields for five year run of a magnetized ICAL detector with the benchmark set of 
active-sterile mixing angles given in Set-1 of Table~\ref{1}. 
We consider the energy of injected muon to be 50 GeV at the muon storage ring directed towards the ICAL detector. 
The estimated numbers are shown for two values of chosen $\Delta m_{41}^2=1\times 10^{-3}$ eV$^2$ 
and $\Delta m_{41}^2=3\times 10^{-3}$ eV$^2$.
As we have discussed before, keeping  $\Delta f_{31}=5\times 10^{-27}$ fixed, 
we compute the right sign and wrong sign muon yields at the ICAL detector considered here 
for different $\Delta f_{41}$ values 
discussed in previous section. 
From Table~\ref{2}, we observe that for both the 
values of $\Delta m_{41}^2=1(3)\times 10^{-3}$ eV$^2$, 
increase in $\Delta f_{41}$ results in depletion of the right sign muon yields while wrong sign 
muon yields do not suffer any significant change.
 %However, for the choice $\Delta m_{41}^2=3\times 10^{-3}$ eV$^2$, although right 
%sign muon yield depletes, no significant change is observed for wrong sign muon yield. 
We also compute the right and wrong sign muon yields for a set of different benchmark 
points for active-sterile mixing given in Set 2 of Table 1 keeping the other parameters same as those used in Table 2. 
The results are furnished in Table 3. Similar trend as in Table 2 is also observed for the computed 
right and wrong sign muon yield values given in Table 3. From these calculations it appears that long 
baseline neutrino experiment can be very effective and important for not only to address,
in addition to other neutrino physics issues, the four neutrino flavour 
scenario but also to probe a possible tiny violation of equivalence principle, if any.
\begin{table}[]
\centering
\caption{Comparison of the right sign and the wrong sign $\mu$ yields for the 
4-flavour case compared to the same for 3-flavour case for two sets of 
active-sterile neutrino mixing angles and for two different values of 
$\Delta m_{41}^2$. We consider $\Delta f_{41} = 0$ for this comparison and muon 
injection energy is kept fixed at 50 GeV.
See text for details.}
\vskip 2mm
\begin{tabular}{|c|c|c|c|c|c|c|c|}
\hline
$\Delta m_{41}^2$ & $\theta_{14}$ & $\theta_{24}$ & $\theta_{34}$ & 
Right & Wrong & Right & Wrong  \\
in eV$^2$ &&& & sign $\mu$ & sign $\mu$ & sign $\mu$ & sign $\mu$ \\
&&& & in  & in & in & in  \\
&&& & 4-flavour & 4-flavour & 3-flavour & 3-flavour \\
\hline
\multirow{2}{*}{$1\times 10^{-3}$} & 3.6$^{\circ}$ & 4.0$^{\circ}$ & 18.48$^{\circ}$  
& 3115192 & 4182 & 2250268 & 463\\ 
& 2.5$^{\circ}$ & 10.0$^{\circ}$ & 30.0$^{\circ}$ & 3386441 & 4501 & 
2250268 & 463 \\ 
\hline
\multirow{2}{*}{$3\times 10^{-3}$} & 3.6$^{\circ}$ & 4.0$^{\circ}$ & 18.48$^{\circ}$ 
& 3006510 & 4138 & 2250268 & 463 \\ 
& 2.5$^{\circ}$ & 10.0$^{\circ}$ & 30.0$^{\circ}$ & 3006875 & 4341
& 2250268 & 463 \\\hline 
\end{tabular}
\label{4}
\end{table}

In order to demonstrate, how both the right sign and wrong sign muon
yields for 4-flavour scenario differ from those when only three 
active flavours are considered with and without VEP effects, 
we compare in Table 4 the yields for the two cases by computing 
the right sign and wrong sign muon yields when three flavour mixing
parameters are same for both the scenarios. The other parameters 
of active-sterile mixing are adopted as given in Sets I and II 
of Table I (and also shown in Table 4). The value of 
$\Delta f_{31} = 5 \times 10^{-27}$, as in Tables 2 and 3 and fixed 
from the bounds on $\Delta f_{31}$ given in Ref. \cite{Esmaili:2014ota}  and 
$\Delta f_{41} =0$. All computations are for five year run of 
the chosen 50kTon ICAL detector assumed to be placed at a distance 
of 7359 km from a neutrino factory with muon injection energy of 50 GeV.

From Table 4, it is clear that for both the cases of 
right sign and wrong sign muon yields are enhanced 
for 4-flavour (3 active + 1 sterlile) scenario when compared with
3-flavour oscillation. 
%scenario is an order of magnitude lower than those with 4-flavour (3 active + 1 sterlile) considerations. 
The difference is more striking for 
the case of wrong sign muon yield. For example, when 
$\Delta m^2_{41} = 3 \times 10^{-3}$ is chosen, the wrong sign muon yield
for 4-flavour case is 20 times larger in magnitude than the
same for 3-flavour case. If the active-sterile 
VEP oscillation is made non-zero ($\Delta f_{41} \neq 0$), the 4-flavour
results will be modified as seen from Tables 2 and 3 but the similar
trend in difference of %right and 
wrong sign muon yields are maintained.
However, for increasing $\Delta f_{41}$, right sign muon yield reduces
considerably and  can become even smaller than the 3-flavour case
as observed in Table 3 for $\Delta f_{41}=10^{-23}$. 
From the estimated results of muon yields obtained in Table~\ref{2}-\ref{3} at the chosen 
ICAL detector, it is now evident that the presence of gravity induced four neutrino oscillations in matter will 
significantly affect the right sign and wrong sign muon yields for non zero $\Delta f_{41}$. Therefore, 
long baseline neutrino experiment can be a viable probe to investigate the 
violation of equivalence principle appearing in four flavour scenario. 

\begin{figure}[h!]
\centering
\subfigure[Variation of right sign $\mu$ yield ]{
\includegraphics[width=7 cm,scale= 0.5,angle=0]{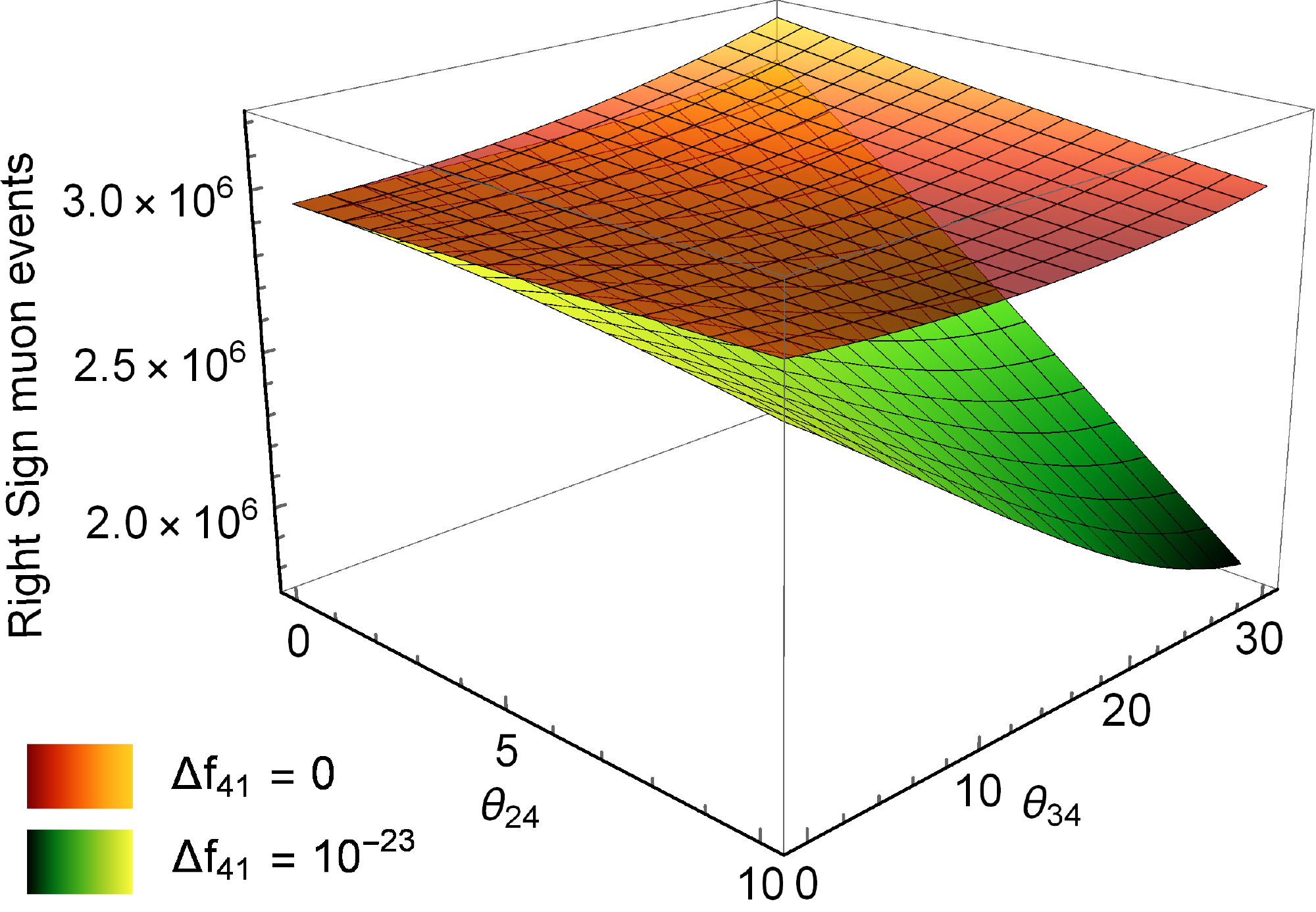}}
\subfigure [Variation of wrong sign $\mu$ yield]{
\includegraphics[width=7 cm,scale= 0.5,angle=0]{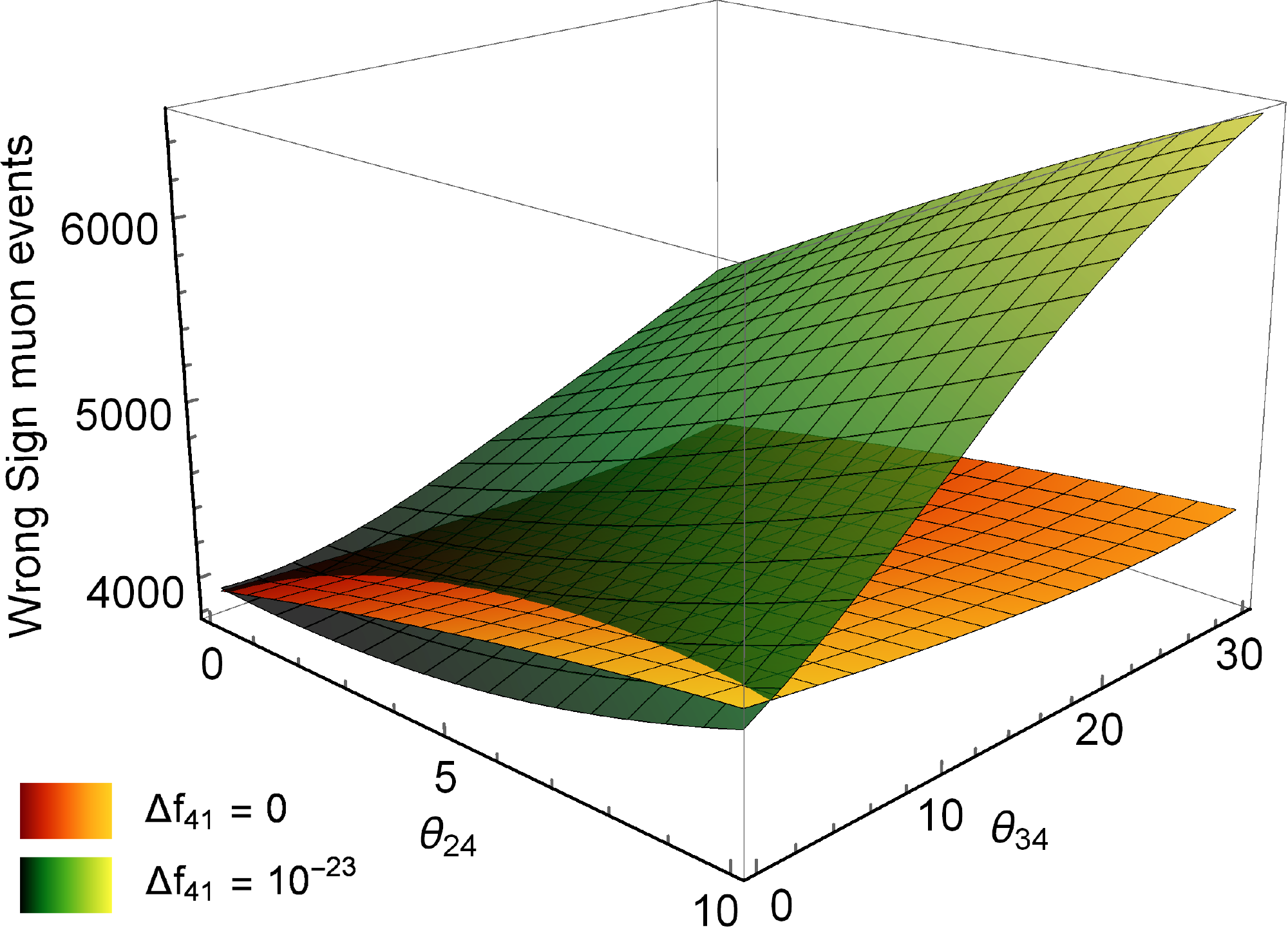}}
\caption{Right sign and wrong sign muon yield variations with $\theta_{24}$ and 
$\theta_{34}$ for fixed $\theta_{14}=2.5^0$, $\Delta f_{31}=5\times 10^{-27}$ 
and two chosen values of $\Delta f_{41}=0,10^{-23}$. See text for details.}
\label{fig6}
\end{figure}

Finally the 3D plots in Fig.~\ref{fig6}, we show the 
simultaneous variations of each of the right sign and wrong sign 
muon yields with $\theta_{24}$ (ranging from $0^0\leq\theta_{24}\leq10^0$) 
and $\theta_{34}$ (ranging from $0^0\leq\theta_{34}\leq30^0$) for fixed values of $\theta_{14}=2.5^0$ 
and $\Delta f_{31}=5\times 10^{-27}$. The chosen ranges of $\theta_{24}$ and $\theta_{34}$ are within the 
allowed regions given by 4-flavour analyses of oscillation experiment data. 
We compute the 
muon yields with muon injection energy 50 GeV in proposed ICAL detector with baseline
length 7359 km for five year run and compare the muon yield values for $\Delta 
m_{41}^2=10^{-3}$ eV$^2$, $\Delta f_{41}=0$ with $\Delta f_{41}=10^{-23}$. 
In Fig.~\ref{fig6}a, we notice that for $\Delta f_{41} = 0$, right sign muon
yield does not change significantly with the variations of 
$\theta_{24}$ and $\theta_{34}$. However, with $\Delta f_{41} = 10^{-23}$, 
one observes considerable 
depletion in right sign muon yield as $\theta_{34}$ increases while the 
changes are negligible with the variations of $\theta_{24}$. This result is 
also in agreement with the 
muon yield events reported earlier in Table~\ref{2}-\ref{4}, where a reduction has been 
observed in right sign muon yield for increase in $\Delta f_{41}$.
A similar plot with the same set of parameters
for wrong sign muon yield events at ICAL detector with 50 GeV muon injection energy (and 
five year run) is shown in Fig.~\ref{fig6}b. 
Comparing the values of 
wrong sign muon yields for $\Delta f_{41}=0$ with $\Delta f_{41}=10^{-23}$,
we observe very mild variation in wrong sign muon yields as the values of 
$\theta_{24}$, $\theta_{34}$ are changed.
The nature of this 3D plot is consistent with the results shown in Tables 
2-4 for wrong sign muons. Thus, this may be concluded that in case of  
4-flavour oscillations the VEP, if exists in nature can influence 
considerable effect on muon yields in a long baseline neutrino experiment.
\section{Summary and Discussions}
If the equivalence principle is indeed violated in nature, this will induce 
different gravitational couplings for different types of neutrinos. Here in 
this work,
we study how such a probability affects the neutrino oscillations in matter and the
possibility that even a very small violation of equivalence principle can be probed by a long baseline
neutrino experiment. To this end we consider a four neutrino (3+1) framework, 
where one extra 
sterile  neutrino is assumed to exist in addition to the three active neutrinos. 
In a possible scenario that the equivalence principle is violated in nature, 
the three active neutrinos as well as the sterile neutrino couple differently
with gravity which result in a gravity induced oscillations of neutrinos in addition to mass flavour oscillation of
neutrinos.
{%As eigenstates in gravity basis are not the same as those in flavour basis,
%neutrino oscillation phenomenology will be significant to study the effects of violation of equivalence
%principle since in such a scenario neutrino mass eigenstates are different from flavour and gravity eigenstates.}
In addition, one must also take into account the mater effects if neutrinos 
propagate through medium.
In this work, we derive the effective Hamiltonian for four neutrino oscillations in presence of
gravity induced effects along with usual mass-flavour oscillations with matter effects.
we then study the effects of VEP in a long baseline neutrino oscillation 
experiment, by estimating the detector yields at the far detector.

We derive the new oscillation probabilities for neutrino oscillations within 
matter assuming the mixing angles
between mass and flavour eigenstates to be identical with those between gravity and flavour eigenstates.
The above choice allows us to study the effects of new parameters $\Delta f_{31}$, $\Delta f_{41}$ which
are responsible for gravity induced neutrino oscillations and can be attributed to the signature and measure of 
violation of equivalence principle. However, IceCube data of atmospheric 
neutrino 
puts stringent constraint on $\Delta f_{31}$. Using the bounds on neutrino mixing angles from different 
experiments and $\Delta f_{31}$, we study the behaviour of four neutrino
oscillation probabilities
considering a representative long baseline of 7000 km for different values of $\Delta f_{41}$
while the values of $\Delta m_{41}^2$  and $\Delta f_{31}$ are kept fixed. For 
demonstrative purpose, we consider two benchmark sets of active-sterile neutrino 
mixing angles $\theta_{14},~\theta_{24},~\theta_{34}$.
We observe that significant deviations in oscillation probabilities $P_{\nu_\alpha \rightarrow \nu_\beta}$
occur with the changes in $\Delta f_{41}$ indicating that even a very weak 
violation of the equivalence principle will affect the oscillation probabilities over a chosen 
representative baseline of 7000 km.

With oscillation probabilities that we derive in this work for 3 active and 1 sterile neutrino formalism,
 we make an estimate of the signatures of  violation of equivalence principle
at the end detector of an assumed LBL neutrino experiment where the neutrinos are 
produced in a neutrino factory and are detected at a far detector of magnetized 
iron calorimeter (ICAL) with a baseline length of around 7359 km.
The magnetized ICAL detector
can efficiently measure the number of $\mu^-$ and $\mu^+$ produced upon charged
current interaction of muon neutrinos and muon anti-neutrinos at the detector. Flux of muon neutrinos
($\nu_\mu$) will suffer gravity induced and mass induced 
oscillations in matter while propagating to the far ICAL detector from the
neutrino factory and  thus it will be depleted. Hence this oscillation channel 
is referred to as disappearance channel.
Similarly detection of $\mu^+$ at magnetized ICAL indicates the appearance 
channel due to neutrino
oscillation $\bar{\nu}_{e}\rightarrow \bar{\nu}_{\mu}$. 
Our study reveals that, since $\Delta f_{31}\simeq 10^{-27}$, VEP 
effect is negligible in case of three flavour neutrino oscillations and also 
negligible in case of four flavour oscillations when $\Delta f_{41}=0$ is taken 
into account. 
From the calculation of various oscillation probabilities, we  
demonstrate that effect 
of VEP becomes prominent with four flavour oscillations for $\Delta f_{41}=10^{-23}$. 
This is further justified by the calculation of muon yields ($\mu^\mp$) in a proposed
long baseline neutrino experiment with baseline length 7359 km. The results show significant
changes in muon yields arising due to VEP parameter $\Delta f_{41}$ responsible for gravity 
induced neutrino oscillations. 
In case of four flavour oscillations, we conclude that the 
effect of VEP is distinguished significantly for $\Delta f_{41} = 10^{-23}$.
%and most prominent for $\Delta f_{41}\sim 10^{-23}$. 
In fact, effect of gravity induced oscillations
of neutrinos in four flavour scenario is determined entirely by the parameter $\Delta f_{41}$.
%Comparing these results with the case when the gravity induced oscillation is absent 
%(i.e. $\Delta f_{31,41}=0$; normal mass induced oscillation) 
%we conclude that for certain choices of $\Delta f_{31}$ and $\Delta f_{41}$, 
%yields of wrong sign muon increase considrably. Similarly for
%some different values of $\Delta f_{31}$ and $\Delta f_{41}$ values, right sign
%muon yields deplete singificantly which could be positive  indications of VEP and gravity induced
%oscillation of neutrinos. 
Therefore, long baseline neutrino experiment can be used to probe even a very 
small violation of equivalence principle, if exists in nature. 
However, non observation of any such deviations
in predicted muon yield will certainly rule out the possibility of the effect of 
VEP in neutrino oscillations.
It is to be noted that one can also perform a detailed study of VEP effect 
where sterile neutrino is considered and put limit on $\Delta f_{41}$, similar to the work  done in Ref.~\cite{Esmaili:2014ota}
to constrain $\Delta f_{31}$. However,  such an analysis is beyond the scope of present work and can be 
pursued in a future work.

{\bf Acknowledgements :} 
Authors acknowledge A. Bandyopadhyay for useful discussions.
One of the authors (M.P.) thanks the DST-INSPIRE 
fellowship (DST/INSPIRE Fellowship/2016/IF160004) grant by Department of Science 
and Technology (DST), Govt. of India. One of the authors (A.H.) acknowledges 
the support received from St. Xavier's College, Kolkata Central Research 
Facility and also thanks the University Grant Commission (UGC) of the 
Government of India, for providing financial support, in the form of 
UGC-CSIR NET-JRF. ADB acknowledges the support and hospitality of SINP for
 the completion of the work. Work of ADB is supported in part by the National 
 Science Foundation of China (11422545,11947235).

\end{document}